\documentclass{revtex4}
\usepackage{latexsym}
\usepackage{epsfig}

\usepackage{amsmath}
\usepackage{amsfonts}
\usepackage{amssymb}
\usepackage{graphicx}

\newcommand{\be}{\begin{equation}}
\newcommand{\ee}{\end{equation}}
\newcommand{\bea}{\begin{eqnarray}}
\newcommand{\eea}{\end{eqnarray}}

\def\pd{\partial}
\def\d{{\delta}}

\def\mc{m_\chi}
\def\mPl{m_{Pl}}
\def\P{{\cal P}}

\def\fnl{{f_{\rm NL}}}
\def\gnl{{g_{\rm NL}}}
\def\rT{{r_T}}
\def\osc{{\rm osc}}
\def\O{\Omega}
\def\p{\phi}
\def\x{\chi}

\def\z{\zeta}
\def\zr{{\zeta_\gamma}}
\def\zc{{\zeta_\x}}
\def\zp{{\zeta_\p}}
\def\e{\epsilon}
\def\rc{{R_\x}}
\def\tnl{{\tau_{\rm NL}}}
\def\wc{w_\x}
\def\gc{{\Gamma_\x}}
\def\Pz{{\P_\z}}

\def\pfw{p_{FW}}
\def\S{{\cal S}}
\def\Sx{{\S_\x}}
\def\Sg{\S_G}

\begin{document}

\title{Primordial non-Gaussianity from mixed inflaton-curvaton perturbations}

\author{Jos\'e Fonseca \& David Wands}

\affiliation{Institute of Cosmology \& Gravitation, University of Portsmouth, Dennis Sciama Building, Burnaby Road,
Portsmouth, PO1~3FX, United Kingdom}

\begin{abstract}
We characterise the primordial perturbations produced due to both inflaton and curvaton fluctuations in models where the curvaton has a quadratic, cosine or hyperbolic potential, and the inflaton potential is characterised by the usual slow-roll parameters. Isocurvature curvaton field perturbations can produce significant non-Gaussianity in the primordial density field, in contrast with adiabatic inflaton field perturbations which produce negligible non-Gaussianity for canonical scalar fields. A non-self-interacting curvaton with quadratic potential produces a local-type non-Gaussianity that is well described by the non-linearity parameter $f_{\rm NL}$, which may be scale-dependent when the inflaton perturbations dominate the power spectrum. We show how observational bounds on non-linearity parameters and the tensor-scalar ratio can be used to constrain curvaton and inflaton parameters. We find a consistency relation between the bispectrum and trispectrum parameters in a mixed inflaton-curvaton model for a quadratic curvaton potential. Self-interaction terms in the curvaton potential can lead to both a large trispectrum parameter, $g_{\rm NL}$, and scale-dependence of the non-linearity parameters.
\end{abstract}

\maketitle

\date{\today}

\section{Introduction}

Inflation is our most successful theory for explaining the initial conditions required for the hot Big Bang cosmology. In particular, primordial density perturbations can be produced from initial quantum fluctuations that are stretched by the accelerated expansion up to super-Hubble scales to become the large-scale structure of the Universe today. This mechanism can give rise to an almost scale-invariant power spectrum, as observed in the cosmic microwave background \cite{Komatsu:2010fb}.

The standard model of inflation is described by a light scalar field $\phi$, the inflaton, whose slow-roll controls the potential energy that drives the accelerated expansion. When inflation ends the inflaton potential energy rapidly decays into radiation. Any light scalar field during slow-roll inflation will acquire a nearly scale-invariant spectrum of perturbations at Hubble exit, $k=aH$, and any such field can, in principle, source structure in the Universe. The curvaton, $\x$, is a light, weakly-coupled field during inflation whose energy density is negligible during inflation, but if the field remains weakly coupled at the end of inflation, its energy density can grow relative to radiation after inflation, and perturbations in the curvaton field can lead to primordial density perturbations when the curvaton decays into radiation \cite{Mollerach:1989hu,Linde:1996gt,Enqvist:2001zp,Lyth:2001nq,Moroi:2001ct,Moroi:2002rd,Lyth:2002my}. A distinctive feature of the curvaton model is that it can leave behind significant local-type non-Gaussianity in the primordial density distribution, $|\fnl|\gg1$.

Previous studies \cite{Nakayama:2009ce,Fonseca:2011iz} have used primordial non-Gaussianity, characterised by the non-linearity parameter $\fnl$, and the tensor-to-scalar ratio, $\rT$, to constrain curvaton model parameters.
In the simplest model of a curvaton with a quadratic potential, the amplitude of primordial density perturbations, together with a measurement of $\fnl$ and $\rT$ would fix the energy scale of inflation, the initial curvaton vacuum expectation value (VEV) and the dimensionless curvaton decay rate. A self-interacting curvaton would introduce additional model parameters that could be fixed by scale-dependent $\fnl$ and higher-order correlators.
These studies focussed solely on density perturbations coming from the curvaton field. But if inflation is driven by a slow-roll inflaton, then there are inevitably fluctuations in the inflaton field too which lead to some level of density perturbations when inflation ends and the inflaton energy decays into radiation. Inflaton perturbations are adiabatic and thus lead to constant curvature perturbations, $\zeta$, on super-Hubble scales, whose local-type non-Gaussianity is always small, $|\fnl|<{\cal O}(1)$.

In this paper we also consider the possibility that both fields contribute for the primordial density perturbation. In particular we note the possibility that while the inflaton contribution to the primordial power spectrum may dominate over that from the curvaton, the curvaton can nonetheless source significant non-Gaussianity.
%
% Several authors have previously considered mixed inflaton-curvaton scenarios.
%
Bartolo and Liddle \cite{Bartolo:2002vf} were the first to consider in what regime the linear curvaton or inflaton field perturbations would dominate the primordial power spectrum in a simple model of two massive scalar fields. Ichikawa et al \cite{Ichikawa:2008iq} also considered non-Gaussianity and the tensor-scalar ratio in several classes of inflation models, including chaotic, hybrid, and new inflation.
In this paper we will evaluate the relative contribution of curvaton field fluctuations to the primordial power spectrum for a general inflaton potential. The contribution of adiabatic inflaton perturbations is given relative to the tensor power spectrum by the slow-roll parameter, $\epsilon_*$. Here we define an analogous parameter, $\e_c$, describing the contribution of curvaton perturbations relative to the tensor power spectrum. The second inflaton slow-roll parameter, $\eta$, then affects only the scale-dependence of the power spectrum.
Very recently Kobayashi and Takahashi have considered the scale-dependence of both the power spectrum and the non-Gaussianity including mixed inflaton-curvaton models \cite{Kobayashi:2012ba}.
Note that Langlois and Vernizzi \cite{Langlois:2004nn} studied linear adiabatic and isocurvature density perturbations allowing for a general inflaton potential, which was then extended \cite{Langlois:2008vk} to second- and higher-order to study non-Gaussianity in both adiabatic and residual isocurvature density perturbations after inflation and their correlation.
%New 24/4 
Recently Kinney et al \cite{Kinney:2012ik} considered the complementary constraints coming from the shape of a mixed (bimodal) power spectrum for adiabatic density perturbations.
In the following we assume that all species are in thermal equilibrium after the curvaton decays and thus there are no residual isocurvature modes \cite{Lyth:2003ip,Weinberg:2004kf}.

% where the inflaton is the source of adiabatic density fluctuations while the curvaton gives rise to residual isocurvature perturbations.
% Note that we no longer require slow-roll inflation. Lyth and Dimopoulos [CITE] studied inflaton models that are allowed in the curvaton scenario.

In Section II we will briefly review the origin of density perturbations coming from both inflaton and curvaton field perturbations, and the spectrum of tensor metric perturbations (gravitational waves). We will review both the linear transfer of field perturbations into radiation, and the non-linear transfer for curvaton perturbations at second- and third-order in the field perturbations which can give rise to a non-vanishing primordial bispectrum and trispectrum.
In Section~III we present our results based on numerical solutions of the curvaton field evolution after inflation \cite{Fonseca:2011iz} and previous numerical studies of curvaton decay based on a fluid description of the curvaton at late times \cite{Malik:2002jb,Gupta:2003jc,Malik:2006pm,astro-ph/0607627}. We focus primarily on the simplest curvaton model with a quadratic potential with a fixed mass, $\mc$. In this case significant non-Gaussianity arises only when the curvaton is sub-dominant when it decays, $\rc\ll1$ and we derive a consistency relation between the bispectrum and trispectrum parameters which holds even in the mixed inflaton+curvaton case. The third-order non-linearity parameter, $\gnl$, remains small, $\gnl\ll\fnl^2$, even in the mixed scenarios. Scale-dependence of $\fnl$ may distinguish the mixed inflaton+curvaton model from the curvaton limit for a quadratic curvaton potential.
We also examine self-interacting curvaton models, including a cosine potential, which introduce an additional mass scale, $f$, where self-interaction terms become important for $\x_*\sim f$. Self-interactions can produce significant non-Gaussianity even if the curvaton dominates when it decays, $\rc\simeq1$. Self-interacting curvatons produce large third-order non-linearity parameter, $\gnl$, as well as scale-dependent $\fnl$. We discuss our results and conclude in Section~IV.

\section{Density perturbations from inflation}

%In this paper we consider that during inflation there exists a scalar field $\x$, the curvaton. It doesn't at any point drive inflation and is completely subdominant during inflation, $\O_\x\ll1$. It is a weakly coupled late-decaying field. The setup we will consider is the following: both the inflaton and the curvaton acquire field perturbations during inflation; inflation terminates with the complete decay of the inflaton into radiation; the curvaton is in an over-damped regime throughout inflation until it starts decaying into radiation during radiation era. We should note the energy hierarchy $H_*>\mc>\gc$, where $H_*$ is the Hubble when the modes leave the horizon ($k=aH$), $\mc^2\equiv d^2V/d\x^2$ is the curvaton mass and $\gc$ is the curvaton decay rate. %Throughout these notes I will keep the notation of \cite{Fonseca:2011iz}.

%[??? I WANT TO CONSIDER THE POSSIBILITY OF NON-SINGLE FIELD INFLATION AND NON SLOW-ROLL. HOW TO INCLUDE THIS IN THE REVIEW?]

\subsection{Background evolution}

We take both the inflaton and the curvaton to be in a slow-roll regime during inflation
% , i.e.
% \be
% 3H\dot\x\simeq-V_\x
% \ee
% where $V_A\equiv\pd V/\pd A$. The
but assume that the Friedmann equation is dominated by the inflaton potential energy
\be
H^2\simeq\frac{V(\p)}{3\mPl^2}\,,
\ee
where $\mPl$ is the reduced Planck mass, $\mPl^{-2}=8\pi G_N$.

We define the slow-roll parameters
\be
\epsilon \equiv -\frac{\dot H}{H^2} \,,
\ee
and
\bea
 \epsilon_A &\equiv& \frac{1}{2}\mPl^2\bigg(\frac{V_A}{V}\bigg)^2 \label{epslidef} \\
\eta_{AB} &\equiv& \mPl^2\frac{V_{AB}}{V} \label{etadef} \,,
\eea
where $V_A\equiv\pd V/\pd A$.
Note that in the slow roll approximation $\epsilon\simeq \sum_A\epsilon_{A} \ll 1$ and in the curvaton scenario we assume that $\e_\x\ll\e_\p$ so that $\epsilon\simeq\e_\p$. We also assume the fields are decoupled so that $\eta_{\p\x}=0$.

\subsection{Perturbations during inflation}

% Let's recap field perturbations during inflation. In a spatially flat gauge the field perturbations evolution equation is
% \be
% \ddot{\d\f_I} + 3H\dot{\d\f_I}+\frac{k^2}{a^2}\d\f+\sum_{J}\left[V_{IJ}-\frac1{a^3\mPl^2}\frac d{dt}\left(\frac{a^3\dot{\f_I}\dot{\f_J}}H\right)\right]\d\f_I=0
% \ee
% The last term can be neglected only for the curvaton. For the moment we assume no interactions between the 2 scalars other than gravitational coupling. The quantization of the field perturbations leeds to the result

During inflation, any light scalar fields (with effective mass less than the Hubble scale, $|\eta|<1$) acquire a spectrum of perturbations due to vacuum fluctuations on sub-Hubble scales being stretched up to super-Hubble scales by the accelerated expansion. In particular the curvaton and inflaton field perturbations on spatially-flat hypersurfaces at Hubble exit have a power spectrum
\be
 \label{powerampcurv}
 \P_{\d\p*} \simeq \P_{\d\x*} \simeq \left(\frac {H_*}{2\pi}\right)^2 \,,
\ee
where we neglect slow-roll corrections, including the cross-correlation between inflaton and curvaton perturbations \cite{Byrnes:2006fr}.

Since the inflaton determines the energy density during inflation, inflaton field perturbations on spatially-flat hypersurfaces, $\delta\phi$, correspond to adiabatic curvature perturbations on uniform-density hypersurfaces at Hubble exit, $\zeta_*=-(H\delta\phi/\dot\phi)_*$ and hence we have
\be \label{pzp}
 \P_{\z*} = \P_{\zp} \simeq \frac1{2\mPl^2\e_*}\left(\frac{H_*}{2\pi}\right)^2
\ee
% If we consider a quadratic potential for both the inflaton and the curvaton this implies $\p_*\gg\x_*$.
%
On the other hand curvaton fluctuations are isocurvature field perturbations during inflation, $\epsilon_\chi\ll\epsilon_\phi$, and remain effectively frozen, $\dot\chi\simeq0$, and hence are gauge-invariant during inflation. In particular we can identify curvaton field perturbations on spatially flat hypersurfaces with relative entropy perturbations~\cite{Gordon:2000hv}
\be
 \S_{\chi*} \propto \left( \delta\chi - \frac{\dot\chi}{\dot\phi}\delta\phi \right)_* \simeq \delta\chi_* \,.
 \ee

The power spectrum of free gravitational waves (tensor metric perturbations), like light scalar fields, only depends on the inflation scale at Hubble exit
\be
\P_{T*} = \frac8{\mPl^2}\left(\frac {H_*}{2\pi}\right)^2 \,.
\ee
The tensor-to-scalar ratio is defined by
\be
\rT\equiv\frac{\P_T}{\P_\zeta}
\ee
Both the adiabatic curvature perturbation and the tensor perturbations remain constant on super-Hubble scales, so we have
a tensor-scalar ratio during inflation \cite{Wands:2002bn}
\be
\left( \frac{\P_T}{\P_{\zp}} \right)_* = 16\e_*\,.
\ee
% The radiation density at the beginning of the radiation era is $\rho_\gamma\simeq3\mPl H_*\left(1-\e_*\right)$.

The tensor spectral index is due solely to the variation of the Hubble scale during inflation
\be
n_T\equiv\frac{d\ln \P_T}{d\ln k}=-2\e_*\,.
\ee
However the inflaton field and curvaton field evolve on super-Hubble scales due to their effective mass, and gravitational coupling for the inflaton field, so their spectral tilts are given to leading order by \cite{Wands:2002bn}
\bea
n_\p-1&\equiv&\frac{d\ln\P_\p}{d\ln k}=-6\e_*+2\eta_{\p\p} \label{1nzp}\\
n_\x-1&\equiv&\frac{d\ln\P_\x}{d\ln k}=-2\e_*+2\eta_{\x\x} \label{1nzc}
\eea

% \subsection{Tensor perturbation}

\subsection{End of inflation and after}

At the end of inflation the inflaton decays completely into radiation transferring its curvature perturbation to the radiation, $\zr=\zp$. We assume that reheating or preheating does not alter the power spectrum of the adiabatic density perturbation on large scales, nor does it alter the fluctuations of the curvaton field on large (super-Hubble) scales.
%
%After inflation the curvaton stays as an isocurvature perturbation.  At this stage the density perturbations on spatially flat hypersurfaces are given by the adiabatic inflaton fluctuations, i.e, $\delta N=\zp$.

The curvaton stays in an over-damped regime until the Hubble rate drops to $H\simeq\mc$. At this point the curvaton starts oscillating and behaves like a pressureless matter fluid.
(We will not consider the possibility of the curvaton driving a second period of inflation~\cite{Dimopoulos:2011gb}, i.e., we assume $\x_*< \mPl$.)
Once the curvaton starts evolving like a pressureless fluid we can write its local energy density on uniform-total-density hypersurfaces, $\rho_\x$, in terms of its homogeneous value, $\bar\rho_\x$, and the inhomogeneous entropy perturbation \cite{Lyth:2004gb,Langlois:2008vk}
\be \label{Sx}
\rho_\x=\bar\rho_\x e^{3\left(\zc-\zp\right)}=\bar\rho_\x e^{\S_\x}
\ee
where $\S_\x\equiv3\left(\zc-\zp\right)$ is the non-adiabatic part of the curvaton perturbation. One should note that in the standard curvaton scenario one takes $\zc\gg\zp$, hence $\S_\x\sim3\zc$. In the mixed inflaton-curvaton case this may no longer hold, therefore the quantity to use is $\S_\x$. When the expansion rate drops to $H\sim \mc$ the curvaton starts oscillating in the bottom of its potential, behaving like a pressureless, non-interacting fluid. At later times, but before the curvaton decays, the potential of the curvaton field can be well approximated by a quadratic potential and its time-averaged energy density can be described by
\be
\rho_\x=\frac12 \mc^2|\x^2| \,.
\ee

One can use Eq. (\ref{Sx}) to determine the relation between the entropy perturbations of the curvaton and its field fluctuations during inflation. In the beginning of oscillation we have
\be \label{sxpertdx}
\bar\rho_\x e^{\S_\x}= \frac12 \mc^2\x_{osc}^2
\ee
where $\bar\rho_\x=\mc^2\bar\x_\osc^2/2$. Note that the subscript \emph{osc} stands for beginning of oscillations. Let's define $\bar\x_\osc\equiv g(\x_*)$ where $g$ accounts for non-linear evolution of the field between inflation and oscillations \cite{astro-ph/0607627}. If the curvaton potential is quadratic and we can neglect the self-gravity of the curvaton, we expect linear evolution. On the other hand if it is not quadratic throughout all evolution we need to correct the field perturbations. It is convenient to expand $\x_\osc$ in terms of field perturbations during inflation, $\d\x_*$, i.e.,
\be \label{chiexp}
\x_\osc\simeq g+g'\d\x_*+\frac12g''\d\x_*^2 + \ldots \,,
\ee
where primes denote derivatives with respect to $\x_*$.
%Using the previous result, (\ref{chiexp}), to second order and
Expanding both sides of Eq.~(\ref{sxpertdx}) up to second order we find that the curvaton entropy perturbation is
\be \label{entropycurvpert}
\S_\x=2\frac{g'}g\d\x_*+\left[\frac{g''}g-\left(\frac{g'}g\right)^2\right]\d\x_*^2+\mathcal{O}(\delta\x^3_*)\,.
\ee
One should note that for a non-interacting, isocurvature field, $\d\x_*$ is a Gaussian random field. Therefore we can separate the curvaton entropy perturbation in a Gaussian linear part and second order term as
\be \label{sxsg}
\Sx=\Sg + \frac14\left(\frac{gg''}{g'^2}-1\right)\Sg^2
\ee
where
\be \label{sgdef}
\Sg\equiv2\frac{g'}g\d\x_*\,.
\ee
Hence, using Eqs. (\ref{powerampcurv}) and (\ref{sgdef}), the power spectrum of entropy perturbations in the curvaton is, at leading order, given by
\be \label{psx}
\P_\Sx = \P_{\Sg} = 4\left(\frac{g'}g\right)^2\left(\frac {H_*}{2\pi}\right)^2\,.
\ee

\subsection{Transfer of linear perturbations}

The curvaton decays into radiation when $H\simeq\gc$. We will consider that the curvaton (and the inflaton) decay prior to CDM freeze-out. Therefore we won't consider any residual isocurvature perturbations after curvaton decay \cite{Lyth:2003ip}.

The primordial density perturbation produced by curvaton decay can be estimated using the sudden decay approximation \cite{Lyth:2001nq}. This assumes that the curvaton happens instantaneously on the total-uniform-density hypersurface $H=\Gamma_\x$.  Before the curvaton decays, $\zr=\zp$. Therefore we know that on the sudden decay hypersurface we have
\be
\rho_\gamma=\bar\rho_\gamma e^{4\left(\zp-\z\right)}\,, \qquad  \rho_\x=\bar\rho_\x e^{3\left(\zc-\z\right)} \,.
\ee
At sudden decay we have that the final radiation energy density  $\bar\rho=\rho_\gamma+\rho_\x$. Hence after decay we have
\be \label{eqdensitytotal}
\left(1-\O_\x\right)e^{4\left(\zp-\z\right)}+\O_\x e^{3\left(\zc-\z\right)}=1\,.
\ee

After the decay we have a constant curvature perturbation on super-Hubble scales. Expanding Eq.~(\ref{eqdensitytotal}) to first order, we have
\bea
\z&=&\rc\zc+(1-\rc)\zp \,,\\
&=&\zp +\frac{\rc}{3}\Sx  \, .
\eea
where \cite{Lyth:2001nq,Lyth:2002my}
\be
\rc = \frac{3\O_\x}{4-\O_\x}\bigg|_{dec} \, .
\ee

Since the adiabatic inflaton field perturbations and the isocurvature curvaton field fluctuations (\ref{powerampcurv}) are uncorrelated, the power spectrum of the total primordial curvature perturbations, after curvaton decay, is given by
\be \label{powertotal}
 \P_\z=\P_\zp+\frac{\rc^2}9\P_\Sx \, ,
\ee
% Note that we assume that there are no interactions between the curvaton and the inflaton.
Following Eq.~(\ref{pzp}), and using Eq.~(\ref{psx}), we can write this as
\be \label{ptwithec}
\P_\z = \frac{1}{2\mPl^2} \left(\frac1{\e_*}+\frac1{\e_c}\right) \left(\frac{H_*}{2\pi}\right)^2\,.
\ee
where we define a quantity
\be
 \label{epsilonc}
\e_c \equiv
% \frac92 \frac1{\rc\P_\Sx} \left(\frac{H_*}{2\pi}\right)^2 \,.
 \frac98\left(\frac g{g'\mPl}\right)^2 \frac1{\rc^2}\, .
\ee
The curvaton contribution to the primordial power spectrum corresponds to $(2\mPl^2\e_c)^{-1}(H_*/2\pi)^2$, i.e., $\e_c$ plays the same role for the curvaton contribution to the final power spectrum as $\e_*$ does for $\P_\zp$ in Eq.~(\ref{pzp}).
Thus $\e_c$ marks the critical value of $\epsilon_*$ between inflaton-domination of the primordial power spectrum and curvaton-domination of the power spectrum.
It follows that we can write
\be
 \label{wcwithec}
w_\x \equiv \frac{\e_*}{\e_*+\e_c}\,.
\ee
$w_\x$ can be seen as the function that weighs the curvaton contribution to the final power spectrum.
For $\e_*\gg\e_c$ the curvaton is the dominant contributor to scalar perturbations and $w_\x\simeq 1$. In the opposite regime, $\e_*\ll\e_c$, the inflaton dominates the primordial power spectrum and $w_\x\ll1$.
% \be
% \beta=\frac{\P_\zp}{\rc^2\P_\Sx/9}\,.
% \ee

The spectral index of the primordial power spectrum is then given by
\be \label{1nz}
n_\z-1\equiv\frac{d\ln\P_\z}{d\ln k} =\wc (n_\x-1)+(1-\wc)(n_\p-1)
\ee
% Note that in the curvaton limit we have $\wc=1$.
Substituting the tilts (\ref{1nzp}) and (\ref{1nzc}) for each field in (\ref{1nz}) we have \cite{Wands:2002bn,Langlois:2004nn}
\be \label{nz-1mix}
n_\z-1=-2\e_*+2\eta_{\x\x}\wc+(1-\wc)(-4\e_*+2\eta_{\p\p}) \,.
\ee
% which agrees with \cite{Langlois:2004nn}.
% In non slow roll...

The running of the power spectrum, assuming slow roll inflation and neglecting curvaton-inflaton interactions, is given by
\be \label{az}
\alpha_\z\equiv\frac{dn_\z}{d\ln k}=\wc\alpha_\x+(1-\wc)\alpha_\p+\wc\left(1-\wc\right)\left(n_\x-n_\p\right)^2
\ee
with \cite{Lyth:2009zz,Wands:2003pw}
\bea
\alpha_\p&=&16\e_*\eta_{\p\p}-24\e_*^2-2\xi_\p^2 \label{azp} \\
\alpha_\x&=&4\e_* \left( -2\epsilon_* + \eta_{\p\p} + \eta_{\x\x} \right) \label{azc}\,,
\eea
where
% $\xi_{AB}^2=(\partial^4V/\partial\p_A^3\partial\p_B)/9H^4$.
$\xi_\p^2=(\partial^4V/\partial\p^4)/9H^4$.

%%%%%%%%%%%%%%%%%%%%

\subsection{Non-linearities}

Using the $\d N$ formalism we identify the non-linear curvature perturbation, $\zeta$, with the perturbed expansion up to a final uniform-density hypersurface, $N$, as a function of the local field values on super-Hubble scales during inflation \cite{Lyth:2005fi}
\be
 \zeta = \sum_{A} N_A \delta\varphi_A + \frac12 \sum_{A,B} N_{AB} \delta\varphi_A \delta\varphi_B + \frac16 \sum_{A,B,C} N_{ABC} \delta\varphi_A \delta\varphi_B \delta\varphi_C +\ldots \,,
 \ee
where $N_A\equiv dN/d\p_A$ and $N_{AB}\equiv d^2N/d\p_Ad\p_B$, etc. We define the first non-Gaussianity parameter in terms of the amplitude of quadratic contribution to $\zeta$ relative to the linear terms:
\be \label{deffnl}
\fnl\equiv\frac56\frac{\sum_{AB}N_AN_BN_{AB}}{\left[\sum_{AB}N_AN_B\delta_{AB}\right]^2} \, ,
\ee

If we consider terms in Eq. (\ref{eqdensitytotal}) up to second order we find \cite{Langlois:2008vk}
\bea
\z&=&\rc\zc+(1-\rc)\zp+\frac{\rc(1-\rc)(3+\rc)}2\left(\zc-\zp\right)^2 \\
&=&\zp +\frac{\rc}3\Sx+\frac{\rc(1-\rc)(3+\rc)}{18}\Sx^2 \label{zeta2order}\, .
\eea
Plugging (\ref{sxsg}) into (\ref{zeta2order}) we find
\be \label{41}
\z=\zp +\frac{\rc}3\Sg+\frac{\rc}{18}\left[\frac32\left(1+\frac{gg''}{g'^2}\right)-2\rc-\rc^2\right]\Sg^2\,.
\ee

It is straightforward to see from Eqs. (\ref{pzp}), (\ref{psx}) and (\ref{powertotal}) that
\bea
N_\x&=&\rc\frac {2g'}{3g} \label{nc}\\
N_\p&=&\frac1{\sqrt{2\e_*\mPl^2}} \, .
\eea
We only need to consider the linear terms from the inflaton since $N_{\p\p}\ll (N_\p)^2$ and $N_{\p\x}=0$.
%
% We assume that there is no interaction terms between the inflaton and the curvaton, i.e. $V_{\p\x}=0$, then the cross derivatives of the integrated expansion vanish, i.e., $N_{\p\x}=0$. Eq(\ref{deffnl}) includes derivatives of the integrated expansion with respect to the fields at horizon exit.
%
Then, the first non-Gaussian parameter (\ref{deffnl}) for curvaton+inflaton simplifies to \cite{Langlois:2008vk}
\be \label{fnlmix}
\fnl=\frac56\frac{N_{\x\x}}{N_\x^2}\wc^2 \, .
\ee

It follows directly from Eq.~(\ref{41}) that $\fnl$ is given by
\be \label{fnlsudden}
\fnl=\left[\frac5{4\rc}\left(1+\frac{gg''}{g'^2}\right)-\frac53-\frac56\rc\right]\wc^2\,.
\ee

Note that taking the derivative of Eq.~(\ref{nc}) with respect to $\x_*$ we get
\be \label{ncc}
N_{\x\x}=\frac 23 \rc\left(\frac{g''}g-\frac{g'^2}{g^2}\right)+\rc'\frac {2g'}{3g}\\
\ee
Then, using Eq.~(\ref{fnlmix}) we find the general expression for $\fnl$ in the mixed curvaton-inflaton scenario
\be
\fnl=\left[\frac5{4\rc}\left(1+\frac{g''g}{g'^2}\right)+\frac54\frac{\rc'g/g'-2\rc}{\rc^2}\right]\wc^2 \label{fnlexpr1}\,.
\ee
Comparing this with Eq.~(\ref{fnlsudden}) obtained in the sudden-decay case we have \cite{astro-ph/0607627}
\be
 \label{rcsudden}
\rc'\frac g{g'}=2\rc-\frac43\rc^2-\frac23\rc^3 \,.
\ee
% so we recover Eq.~(\ref{fnlsudden}).

The third order non-linear parameters are $\gnl$ and $\tnl$. They are defined by \cite{Lyth:2005fi,Byrnes:2006vq}
\bea
\gnl&\equiv&\frac{25}{54}\frac{\sum_{ABC}N_AN_BN_CN_{ABC}}{\left[\sum_{AB}N_AN_B\delta_{AB}\right]^3} \label{defgnl}\, ,\\
\tnl&\equiv&\frac{\sum_{ABC}N_AN_BN_{AC}N_{BC}}{\left[\sum_{AB}N_AN_B\delta_{AB}\right]^3} \label{deftnl}\, .
\eea
For the inflaton+curvaton case Eqs.~(\ref{defgnl}) and (\ref{deftnl}) reduce to
\bea
\gnl&=&\frac{25}{54}\frac{N_{\x\x\x}}{N_\x^3}\wc^3 \label{gnlmix} \, , \\
\tnl&=&\frac{N_{\x\x}^2}{N_\x^4}\wc^3 \label{tnlmix} \, .
\eea
Taking the third derivative of $N$ with respect to $\x_*$ we find
\be \label{nccc}
N_{\x\x\x}=\frac23 \rc_{,dec}\left(\frac{g'''}g-3\frac{g''g'}{g^2}+2\frac{g'^3}{g^3}\right)+\frac 43 \rc'_{,dec}\left(\frac{g''}g-\frac{g'^2}{g^2}\right)+\rc''_{,dec}\frac {2g'}{3g} \, .
\ee
Substituting Eqs.~(\ref{nc}) and~(\ref{nccc}) into Eq.~(\ref{gnlmix}) we get
\be
\gnl=\frac{25}{24}\left[\frac1{\rc^2}\left(\frac{g'''g^2}{g'^3}-3\frac{g''g}{g'2}+2\right)+2\frac{\rc'}{\rc^3}\left(\frac{g''g^2}{g'^3}-\frac g{g'}\right)+\frac{\rc''}{\rc^3}\frac{g^2}{g'^2}\right]\wc^3
 \,.
 \label{gnlexpr1}
\ee
Using Eq.~(\ref{rcsudden}) for the sudden-decay approximation we can eliminate the derivatives of $\rc$ to obtain~\cite{astro-ph/0607627}
\be
\gnl = \frac{25}{54} \left[ \frac{9}{4\rc^2} \left(\frac{g'''g^2}{g'^3}+3\frac{g''g}{g'2}\right)-\frac{9}{\rc}\left(1+\frac{g''g}{g'^2}\right)+\frac12\left(1-9\frac{g''g}{g'^2}\right) +10\rc +3\rc^2 \right]\wc^3
 \,.
 \label{gnlsudden}
\ee

Equations~(\ref{fnlexpr1}) and (\ref{gnlexpr1}) do not rely on the sudden decay approximation. Nonetheless Ref.~\cite{astro-ph/0607627} showed that the sudden-decay formulas~(\ref{fnlsudden}) and (\ref{gnlsudden}) do give a good fit to $\fnl(\rc)$ and $\gnl(\rc)$ from the full numerical solution with continuous decay. For example, $\fnl(\rc)$ is accurate to 1\% for $\fnl>60$. Therefore in the following we will use Eq.~(\ref{fnlsudden}) and (\ref{gnlsudden}) to give the non-linearity parameters as a function of $\rc$.

Note that from Eqs.~(\ref{fnlmix}) and~(\ref{tnlmix}) we have
\be
 \label{inequality}
\tnl=\frac{1}{\wc} \left( \frac65 \fnl \right)^2 \, .
\ee
The inequality $\tnl\geq (6\fnl/5)^2$ is an important test of non-Gaussianity the mixed curvaton+inflaton scenario and multi-field scenarios in general \cite{Suyama:2007bg} with equality only in the curvaton limit, $w_\x\to1$.

\section{Numerical results}

We now wish to compute observable quantities such as $\fnl$, $\gnl$ and $\rT$ for different model parameters. We assume that the curvaton is effectively frozen during inflation, which should be a good approximation while the effective curvaton mass is much less than the inflationary Hubble scale, i.e, $\eta_{\x\x}\ll1$. We will then numerically solve for the local evolution of the curvaton field during the radiation- and, possibly, curvaton-dominated epochs after inflation, until the curvaton starts oscillating in the minimum of its potential and behaves like a pressureless fluid, but before it decays. We allow approximately $10^3$ oscillations, i.e., we assume sufficiently slow decay, $\Gamma/m<10^{-3}$, consistent with the hypothesis that the curvaton is weakly coupled to other fields.

We numerically solve the Klein-Gordon equation for $\chi$ prior to decay
\be \label{chievolveq}
\ddot{\x}+3H\dot{\x}+V_\x=0 \,,
\ee
where the Friedmann equation takes the form
\be \label{hp}
H^2=\frac{1}{3\mPl^2} \left(\rho_\gamma+\rho_\x\right) \,,
\ee
and the curvaton density is given by $\rho_\x=\dot\x^2/2+V(\x)$. This allow us to take different potentials for the curvaton field.

The initial conditions to solve Eq. (\ref{chievolveq}) are $\x_i\simeq\x_*$ and $\dot\x_i\simeq -V_\x/3H_i$, since the curvaton is slow-rolling down its potential.
% This will evolve the curvaton from the end of inflation until it starts oscillating. In reality we do not need to care about the details of the end of inflation. We just need to ensure the evolution of the curvaton is over-damped. Therefore we can say that the field values have not evolved much since inflation.
% We take $H_i$ to be the initial Hubble parameter. The Hubble parameter also needs to be evolved.
We take the universe to be radiation dominated initially, $\rho_{\gamma,i}\gg \rho_{\x,i}$ and hence $H_i^2\gg V(\x_*)/3\mPl^2$.
% Nevertheless we will include the of the curvaton in the evolution of the Hubble parameter. Therefore
% The radiation density comes from inflation density. One could say that to first order $\rho_\gamma\simeq\rho_\p\simeq V_{\p\p}(\p)\p_*^2=3H^2_*\eta_{\p\p}\p^2_*$. Therefore, we would need to fix $H_*$, $\eta_{\p\p}$ and $\p_*$ to set initial conditions if re-heating is our starting point. On the other hand we know that $\rho_\gamma \simeq 3H_{end}a^{-4}$ ($a_{end}=1$), where we consider the curvaton completely negligible at the end of inflation. We can approximate the Hubble parameter at the end of inflation and at horizon exit via the relation, $H_{end}\simeq H_*(1-\e_*)$. To set the initial conditions we need to specify two inflation parameters, $H_*$ and $\e_*$. We will take a different approach. We will start solving the equations during the radiation era while the curvaton is still subleading, i.e., $\rho_{\x,i}/\rho_{\gamma,i}\ll1$.
%mainly from the fact that $\rho_{\gamma,i}/\rho_{\gamma,*}\leqslant1$.
In practice we set $\rho_{\gamma,i}=10^{41}\mc^2{\rm GeV}^2$ in our numerical solutions, where $\mc$ is the mass of the curvaton at late times. This ensures that $H_i>100 \mc$ and $\rho_{\gamma,i}\gg \mc^2\x_i^2$ for any $\x_i<\mPl$.

Once the curvaton starts oscillating we can compute the quantity \cite{Fonseca:2011iz}
\be
 \pfw \equiv \Omega_\chi (1-\Omega_\chi)^{-3/4} \sqrt{\frac{H}{m_\x}} \,,
\ee
which becomes constant during oscillations as $\rho_\x\rightarrow {\rm const}/a^3$ and we have $g^2=\chi_{osc}^2\propto \pfw$. 
In this way we connect the scalar field description of the curvaton with a fluid description which has previously been used to numerically study the decay of the curvaton \cite{Malik:2002jb,Gupta:2003jc,Malik:2006pm,astro-ph/0607627}.
Following \cite{Fonseca:2011iz} we can compute the efficiency parameter $\rc$ using the fitting formula \cite{Gupta:2003jc}
\begin{equation}
 \label{Rp}
 \rc(\pfw) \simeq 1 - \left( 1+\frac{0.924}{1.24}\sqrt{\frac \mc\gc}\pfw \right)^{-1.24} \,.
\end{equation}
We find that, for a given curvaton potential, $\pfw$ is dependent only of the initial curvaton field value, $\x_i$. Therefore $\rc$ is dependent only upon the initial curvaton field value and the dimensionless decay rate, $\Gamma_\x/m_\x$.

Going beyond our previous work \cite{Fonseca:2011iz} we will include inflaton perturbations in addition to curvaton field perturbations in our computation of the primordial density perturbation. However the inflaton perturbations represent adiabatic perturbations on super-Hubble scales, i.e., local perturbations along the same background trajectory \cite{Gordon:2000hv}, and they can be treated independently of the curvaton field perturbations.
Looking at the total power spectrum, Eq.~(\ref{powertotal}) we see that we have gained an extra free degree of freedom, $\e_*$, with respect to the purely curvaton limit ($\e_*\gg\e_c$). Therefore our free parameters will be $\gc/\mc$, $\x_*$ and $\e_*$, for the quadratic curvaton \cite{Fonseca:2011iz}. Going beyond the quadratic curvaton potential we will consider models where self-interactions lead to the potential becoming flatter or steeper beyond a characteristic mass scale, $f$, introducing one new parameter in addition to the curvaton mass about the minimum of its potential.

WMAP 7 \cite{Komatsu:2010fb} gives $\P_\z\simeq2.43\times 10^{-9}$ for the amplitude of the power spectrum of curvature perturbations. We will use this observational constraint to fix the inflationary scale. Since the power spectrum (\ref{powertotal}) is proportional to $H_*^2$ we can find a value of $H_*$ that gives the correct power for any values of the other curvaton model parameters. Using Eqs. (\ref{ptwithec}) and (\ref{wcwithec}) we arrive to the constraint equation
\be
H_* = 2\sqrt{2} \pi \mPl \left(\e_c^{-1}+\e_*^{-1} \right)^{-1/2} \Pz^{1/2} \,,
\ee
where the value of $\e_c$ is determined numerically via the formula
\be
\e_c=\frac92\left(\frac \pfw{\pfw'\mPl\rc}\right)^2\,.
\ee
%For a different set of model parameters ($\e_*,\x_*,\mc/\gc$) the constraint equation will give us different values for the Hubble parameter at horizon crossing.
The tensor-to-scalar ratio is then
 \be \label{rt}
 \rT=16w_\x\e_c=16\e_*(1-\wc)\,.
\ee
% Note that $\rT<16\e_*$, since $w_\x>0$.

\subsection{Quadratic potential}

\begin{figure}
\centering
\includegraphics[width=0.6\textwidth]{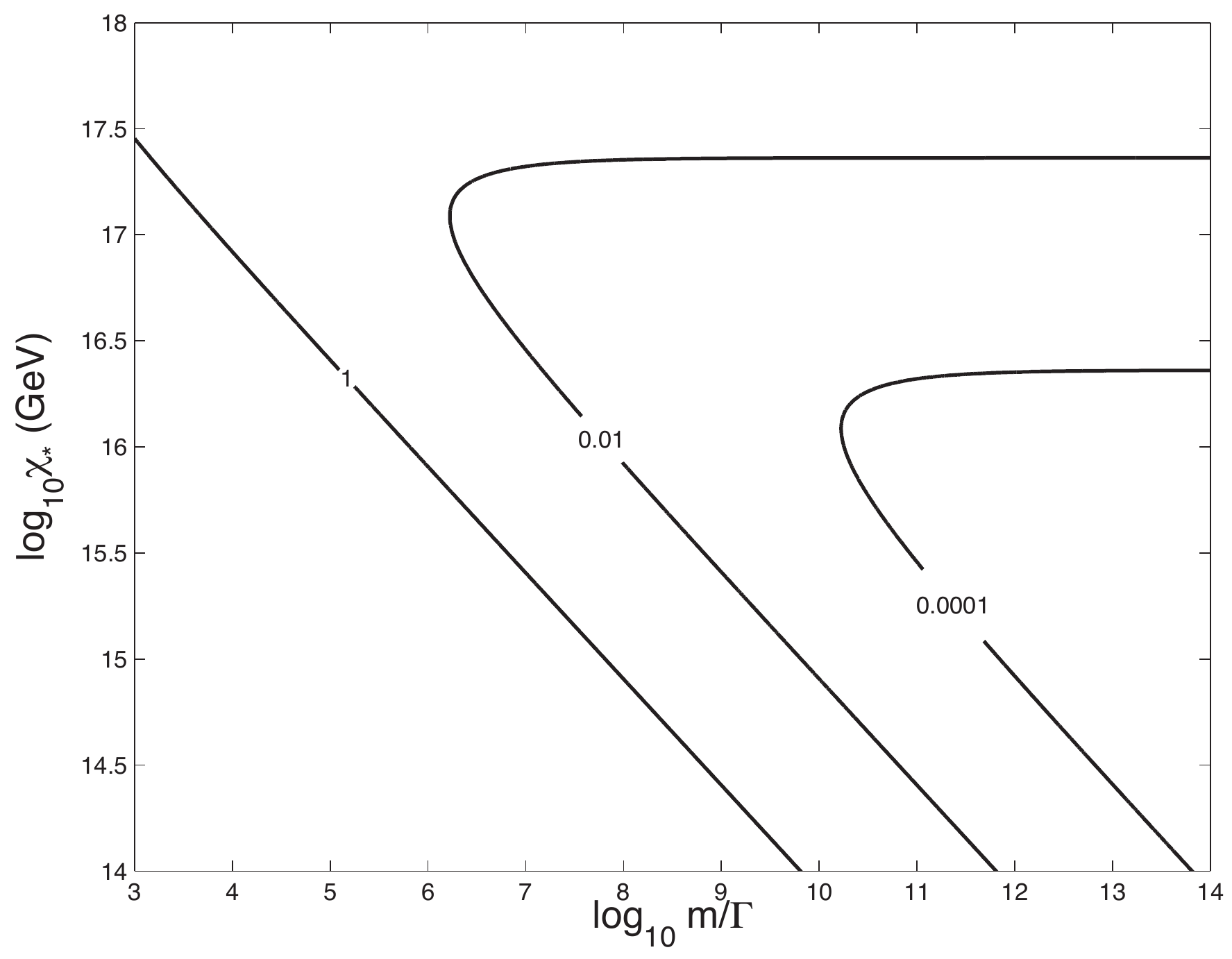}
\caption{The plot show contours of $\e_c$ defined in Eq.~\ref{epsilonc} as a function of the curvaton parameters, $\chi_*$, the curvaton VEV, and $m_\chi/\Gamma_\x$ the dimensionless decay time, for the quadratic curvaton potential, Eq.~(\ref{Vquad}). Curvaton perturbations dominate the primordial scalar power spectrum for $\e_*\gg\e_c$.}
\label{figepscrit}
\end{figure}

The simplest potential consistent with a curvaton scenario is a quadratic potential
\be
 \label{Vquad}
V(\x) = \frac12 \mc^2\x^2 \,,
\ee
and we will see that it is also a good decription of the behavior in more general cases when the curvaton is sufficiently close to the minimum of its potential.

We start by studying the relative contribution of the curvaton to the scalar power spectrum characterised by parameter $\epsilon_c$ defined in Eq.~(\ref{epsilonc}). In Figure~\ref{figepscrit} we plot $\e_c$ as a function of the initial curvaton VEV, $\chi_*$, and the dimensionless decay time $m_\x/\Gamma_\x$.

While the curvaton remains subdominant in the radiation era, we have an analytic solution for the curvaton field \cite{Langlois:2004nn,Ichikawa:2008iq}
and we find \cite{Fonseca:2011iz}
\be
 \pfw \simeq \frac{1.046\chi_*^2}{3\mPl^2} \,.
 \ee
We can clearly identify the two analytic regimes in Figure~\ref{figepscrit}
\be
 \epsilon_c \simeq
 \left\{
 \begin{array}{ll}
 1.125 \left( \frac{\chi_*}{\mPl} \right)^2 & {\rm for}\ \rc \simeq 1 \,,\\
 10.8 \frac{\Gamma_\x}{m_\chi} \left( \frac{\mPl}{\chi_*} \right)^2  & {\rm for}\ \rc \ll 1 \,,
\end{array} \right.
\ee
corresponding to the straight lines in Figure~\ref{figepscrit}.

If we now include the contribution from inflaton perturbations to the total scalar perturbation,
we can identify 3 regimes of interest which depend on the value of $\e_*$ for a given $\epsilon_c$:
\begin{enumerate}

\item The curvaton limit corresponds to $\e_c\ll\e_*$. In this case most of the structure in the universe comes from the curvaton, i.e.,  $w_\x \simeq 1$ from Eq.~(\ref{wcwithec}).
    %Another way of writing this requirement is $\rc^2\P_\zc/9\gg\P_\zp$, or $\beta\ll1$.
    This case has been studied in our previous work \cite{Fonseca:2011iz} and in most of the curvaton literature. We can identify this limit in Figure~\ref{figepscrit} for a fixed value of $\e_*$ as the region inside the contours towards the right of the plot, i.e., for long decay times ($\Gamma_\x\ll m_\x$).

From Eq.~(\ref{rt}) we have in the curvaton limit
\be \label{rtec}
\rT\simeq16\e_c\,.
\ee
Therefore upper bounds on the tensor-to-scalar ratio place constraints on $\e_c$ but do not directly constrain $\e_*$ since $\e_*\gg\e_c$.

In the curvaton limit, $w_\x \simeq 1$, and assuming an effectively massless curvaton, $\eta_{\x\x}\ll1$, then Eq. (\ref{nz-1mix}) gives a red spectral tilt, $n_\z-1\simeq-2\e_*$. In this limit the tilt gives a measurement of the first slow roll parameter, $\e_*$. Consider a fiducial value $n_\z\simeq0.96$ consistent with WMAP7 \cite{Komatsu:2010fb}. For this value of $\e_*\simeq0.02$ we can identify the curvaton limit with the region to the right of the contour $\e_c=0.02$ in Figure~\ref{figepscrit}.

\item The second limit of interest is $\e_c\gg\e_*$. This is the case for which curvaton perturbations are sub-leading in the scalar power spectrum, i.e., $w_\x \ll 1$ in Eq.~(\ref{wcwithec}). These regions correspond to a parameter range where the decay happens too fast (bottom left of the plot), or the curvaton VEV is too big (top) suppressing the curvaton power spectrum. The region $\e_c\gtrsim1$, in Fig. \ref{figepscrit}, will always be in this inflaton dominated limit in slow-roll inflation since $\e_*\ll1$.
%This corresponds to the top and bottom left regions of Figure~\ref{figepscrit}.

The tensor-scalar ratio $\rT$ directly constrains the slow-roll parameter $\epsilon_*$ in the this limit.
% From Eq. (\ref{nz-1mix}) and assuming $\eta_{\p\p}\ll\e_*$ we get $\e_*\simeq0.007$. This gives $\rT\simeq0.1$ which can be detected in future probes.
{}From Eq. (\ref{rt}) we have
\be
\rT \simeq 16\epsilon_* \,.
\ee
The spectral tilt of the primordial scalar power spectrum (\ref{nz-1mix}) is determined by the usual inflaton slow-roll parameters, $n_\zeta-1\simeq-6\epsilon_*+2\eta_{\phi\phi}$ for $w_\chi\ll1$.

In this limit the presence of the curvaton may still be important to as a source of primordial non-Gaussianity or residual isocurvature perturbations after the curvaton decays \cite{Langlois:2008vk}.

\item The third region of parameter space corresponds to $\e_*\sim\e_c$ which corresponds to a mixed scenario.
%If we consider $\eta_{\x\x}\ll\eta_{\p\p}\ll\e_*$ than we expect from Eq. (\ref{nz-1mix}), $\e_*\simeq0.01$. This is of the same order of magnitude of the expected one in the curvaton limit. It predicts $\rT\simeq0.08$ which can be observed in the near future with Planck. Note that the tensor-to-scalar ratio fixes inflation scale, $H_*\simeq10^{15}\rm GeV$.
%
In this case the tensor-scalar ratio (\ref{rt}) no longer directly constrains $\e_*$ or $\e_c$ but the combination
\be
 \label{rTmixed}
 \rT = \frac{16\e_c\e_*}{\e_c+\e_*} \,.
\ee
For example, an observed tensor-scalar ratio, $\rT$, places a lower bound on the slow-roll parameter, $\e_*\geq\rT/16$.

\end{enumerate}

\begin{figure}
\centering
\includegraphics[width=0.6\textwidth]{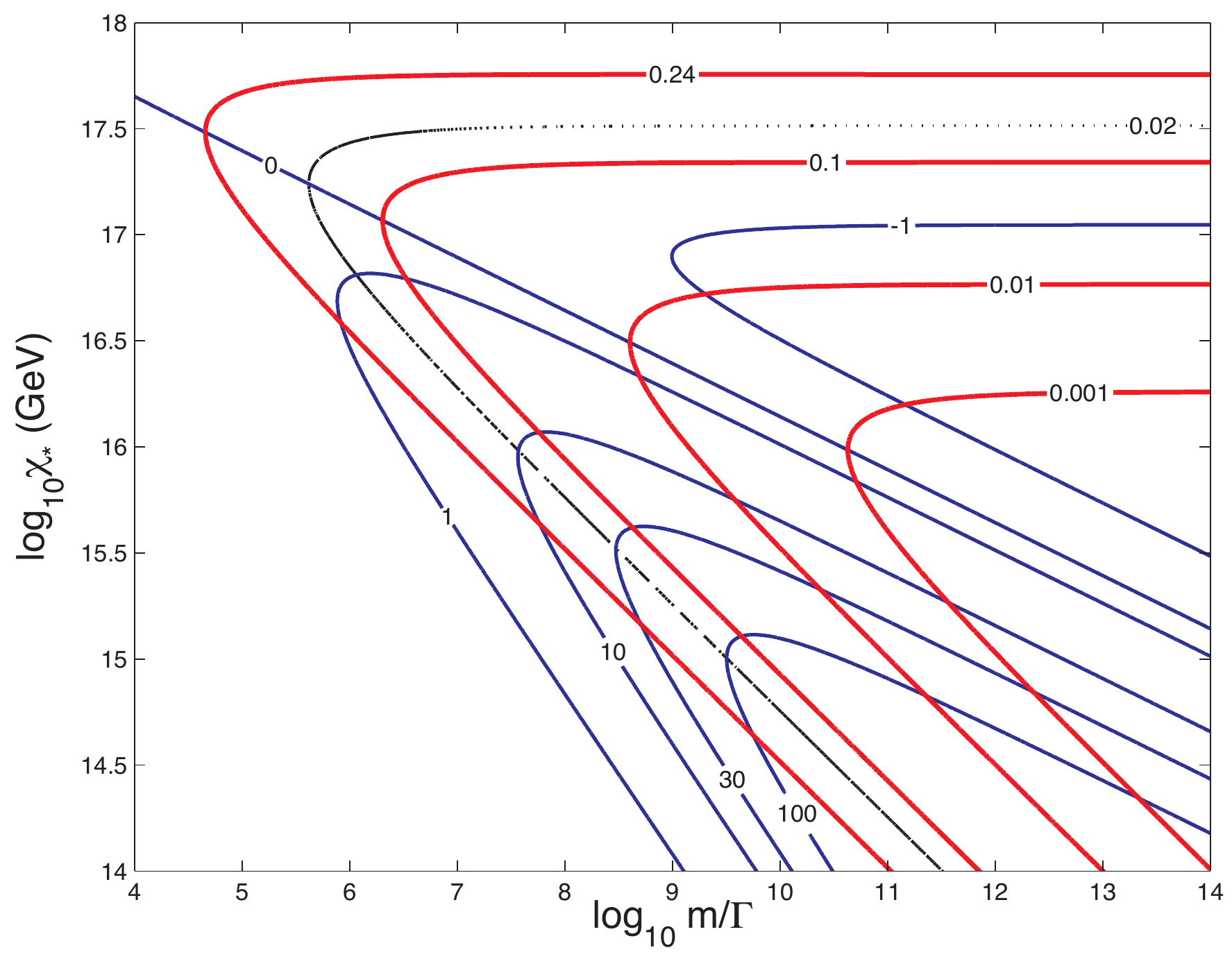}
\caption{The plot show contour lines for the non-linear parameter $\fnl$ (blue lines) and the tensor-to-scalar ratio $\rT$ (thick red lines) as a function of the curvaton parameters, $\chi_*$ and $m_\chi/\Gamma_\x$, for the quadratic curvaton potential, Eq.~(\ref{Vquad}), and a fixed value of the inflation slow-roll parameter, $\e_*=0.02$. The black broken line corresponds to $\e_c=0.02$.
%The primordial scalar power spectrum is dominated by inflaton perturbations above and to the left of the thin black dotted line.
}
\label{figfnlrgweps02}
\end{figure}

In Figure~\ref{figfnlrgweps02} we show contour plots for the non-Gaussianity parameter, $\fnl$, and the tensor-scalar ratio, $\rT$, for the case $\e_*=0.02$. The thin black dotted line is the contour line $\e_c=0.02$ which marks the borderline between the region (1) described above with curvaton-dominated primordial power spectrum and region (2), inflaton-dominated. The curvaton limit, region (1), lies to the right of the $\e_c=0.02$ contour.

We also plot the current observational upper bound on the tensor-scalar ratio, $\rT\lesssim0.24$ \cite{Komatsu:2010fb}. For a given value of $\e_*$, the contours of the tensor-to-scalar ratio follow the contours of $\e_c$ plotted in Figure~\ref{figepscrit}, as expected from Eq.~(\ref{rTmixed}). However, rather then growing without bound as $\e_c$ becomes large, as happens if we consider only the curvaton perturbations~\cite{Fonseca:2011iz}, in the presence of a finite $\e_*$ the tensor-scalar ratio saturates with $\rT\to16\e_*$ in region (2) where $\e_c\gg\e_*$. For $\e_*=0.02$, for example, the tensor-scalar ratio is bounded so that $\rT\leq0.32$.

Similarly the inflaton's (Gaussian) contribution to the primordial scalar power spectrum suppresses the non-linearity parameter $\fnl$ for $\e_c>\e_*$ in region (2). We see that the largest values for $\fnl$ occur in region (3), near the boundary between the curvaton- and inflaton-dominated power spectra, where $\e_c\simeq\e_*$. In the absence of any inflaton perturbations ($w_\chi=1$), the non-Gaussianity continues to grow without bound as $\chi_*/\mPl\to0$ for a fixed value of $m_\chi/\Gamma_\x$~\cite{Fonseca:2011iz}. But $\epsilon_c$ also becomes large as $\chi_*/\mPl\to0$ and therefore the inflaton perturbations dominate the primordial power spectrum. {}From Eq.~(\ref{fnlexpr1}) we see that $\fnl$ is suppressed by an additional factor $w_\chi\simeq\e_*^2/\e_c^2$ and we have
\be
 \fnl \simeq \frac{5}{4\rc} \frac{\e_*^2}{\e_c^2} \simeq 0.038 \e_*^2 \left( \frac{m_\chi}{\Gamma_\x} \right)^{3/2} \left( \frac{\chi_*}{\mPl} \right)^2 \,,
 \ee
which is suppressed as $\chi_*/\mPl\to0$ for a given $\mc/\Gamma_\x$.

If we demand a lower bound on the non-Gaussian parameter, $\fnl>10$ for example, this places an upper bound on the curvaton VEV, $\chi_*<1.2\times10^{16}$~GeV for $\e_*=0.02$, but also a lower bound on the decay rate $\Gamma_\x < 3\times10^{-8}\mc$.

% [I'M CURIOUS ABOUT A LYTH BOUND IN THIS CASE. IT WOULD GIVE A LOWER THEORETICAL BOUND FOR rT].  - I DON'T THINK THERE IS A LYTH-BOUND-TYPE UPPER LIMIT ON THE CURVATON VEV FROM THE TENSOR-SCALAR RATIO, SINCE THE TENSOR-SCALAR RATIO BECOMES INDEPENDENT OF THE CURVATON VEV AS THE VEV BECOMES LARGE (SINCE WE ENTER THE INFLATON DOMINATED REGIME.]

%In Fig. \ref{figfnlrgweps02} we can also find the third region described above. This correspond to the narrow region near $\e_c=0.02$. On its left and towards $\rT=0.24$ we are approaching inflation domination of the primordial power spectrum. For $\e_*=0.02$ this region implies a tensor-to-scalar ratio value which is already excluded from observations. It's also quite noticeable that beyond $\rT=0.24$ we have dominant inflation perturbations since $\fnl$ is strongly suppressed ($w_\x\ll1$).

\begin{figure}
\centering
\includegraphics[width=0.6\textwidth]{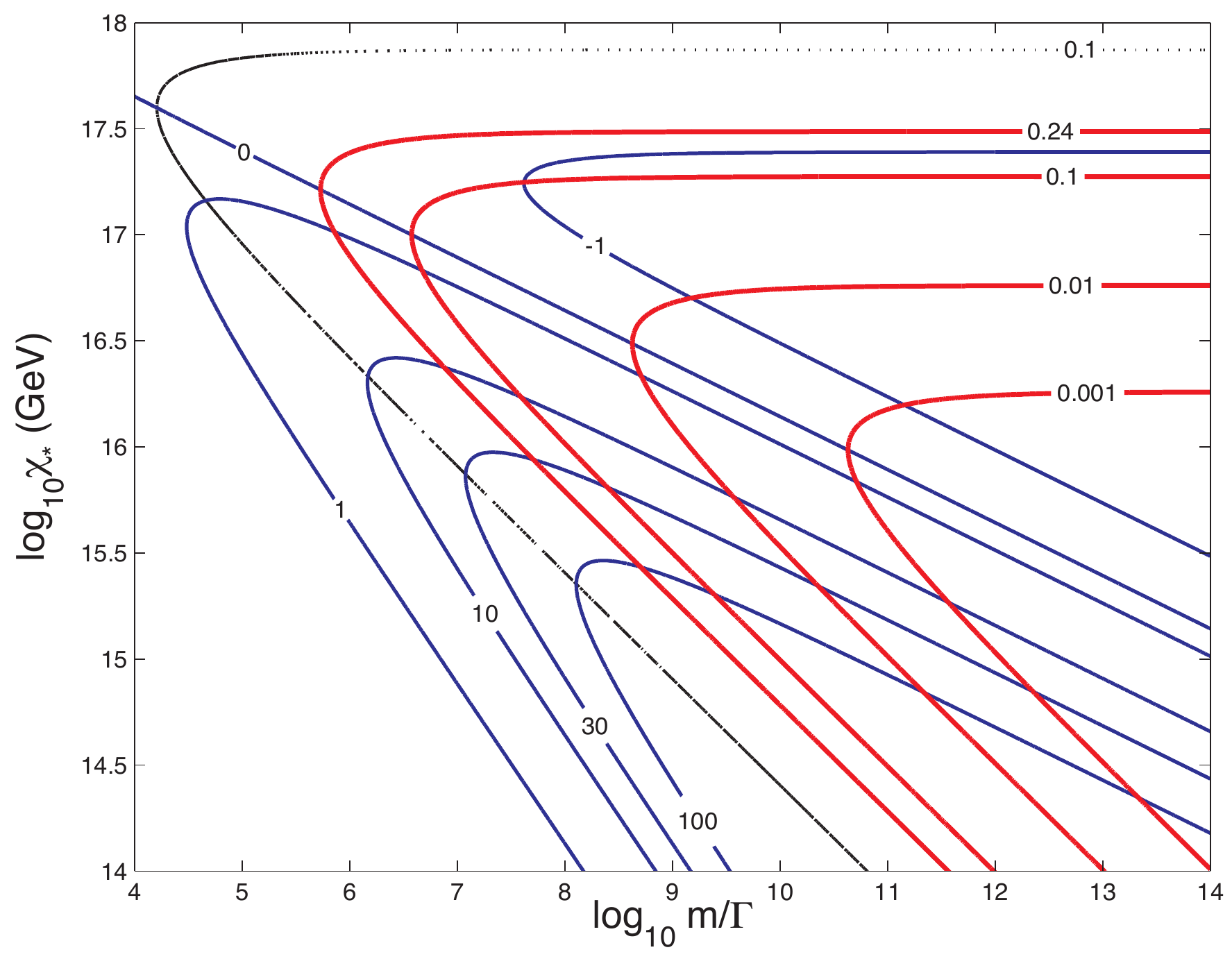}
\caption{The plot show contour lines for the non-linear parameter $\fnl$ (blue lines) and the tensor-to-scalar ratio $\rT$ (thick red lines) as a function of the curvaton parameters, $\chi_*$ and $m_\chi/\Gamma_\x$, for the quadratic curvaton potential, Eq.~(\ref{Vquad}), and a fixed value of the inflation slow-roll parameter, $\e_*=0.1$. The black broken line corresponds to $\e_c=0.1$.}
\label{figfnlrgweps1}
\end{figure}

Figure~\ref{figfnlrgweps1} is similar to Figure~\ref{figfnlrgweps02} but corresponds to a larger slow-roll parameter $\e_*=0.1$. The thin black dotted line separating the inflaton- and curvaton-dominated regions here corresponds to $\e_c=0.1$. We see that for larger values of $\e_*$ the parameter regime (1) corresponding to the curvaton limit extends to smaller values of $m_\chi/\Gamma_\x$ and larger $\chi_*$, permitting larger values of $\fnl$.

On the other hand observational bounds on the tensor-scalar ratio now place more severe constraints on the allowed parameter values.
If we put a lower bound on the non-Gaussian parameter, $\fnl>10$ for example, this places an upper bound on the curvaton VEV, $\chi_*<2.5\times10^{16}$~GeV, and a lower bound on the decay rate, $\Gamma_\x < 2\times10^{-7}m_\chi$, for $\e_*=0.1$ and $\rT<0.24$.

The entire inflaton dominated region (2) is excluded by observational bounds on the tensor-scalar ratio for such a large value of $\e_*$. On the other hand $\e_*=0.1$ is allowed in much of the curvaton dominated region (1).

Note however that such a large value of $\e_*$ requires a similar positive value of $\eta_{\chi\chi}$, tuned such that the spectral tilt remains small in the curvaton limit, $|n_\zeta-1|\simeq 2|\eta_{\chi\chi}-\e_*|<0.1$. Note that inflaton mass, $\eta_{\phi\phi}$, does not affect the spectral tilt in the curvaton limit so the inflaton mass could be of order the Hubble scale without producing a large spectral tilt, but it does affect the running of the spectral index. The running (\ref{az}) in the curvaton limit $w\chi\simeq1$ is $\alpha_\z\simeq\alpha_{\S_\chi}$ which is of the same order of magnitude as the spectral index for $\eta_{\phi\phi}\sim1$.

% By fixing the value of the first slow parameter we also fix the running of the power spectrum. In the curvaton limit $\alpha_\z\simeq\alpha_\Sx\simeq -8\e_*^2=-3.2\times 10^{-3}$. In this limit the running is one order of magnitude lower that the tilt. Note that in the regime $w_\x\simeq1$ and $ \e_*\gg\eta_{\p\p}\gg\eta_{\x\x}$, it follows from Eqs. (\ref{nz-1mix}) and (\ref{azc}) the consistency relation
% \be
% \alpha_\z\simeq-2\left(n_\z-1\right)^2\,.
% \ee

\subsection{Self-interacting potential}

\subsubsection{Cosine potential}

We consider an axion-type potential for a weakly broken $U(1)$-symmetry ($f\gg M$) \cite{Lyth:2001nq,Dimopoulos:2003az,Kawasaki:2008mc,Chingangbam:2009xi,Huang:2010cy}
\be
 \label{Vcos}
 V(\chi)= M^4 \left[ 1- \cos\left( \frac{\chi}{f} \right) \right] \,.
 \ee
For $\chi_*\ll f$ the effective potential  reduces to the quadratic potential (\ref{Vquad}) with $\mc\equiv M^2/f$, but the cosine potential has self-interaction terms which become significant for $\chi_*\sim f$. By symmetry it is enough to consider the range $0\leq\chi_*/f\leq\pi$ for the curvaton VEV during inflation.

\begin{figure}
\centering
\includegraphics[width=0.6\textwidth]{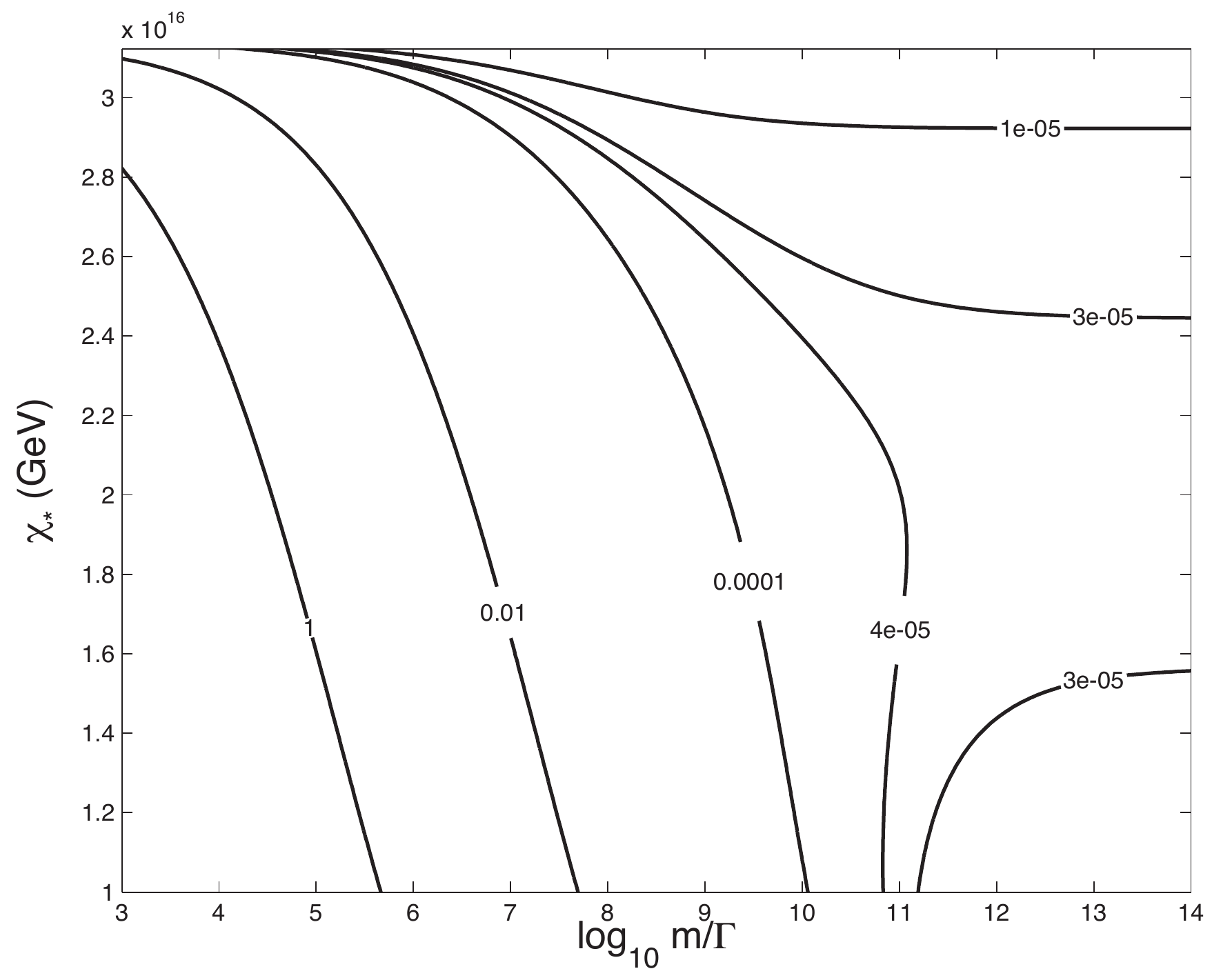}
\caption{The plot show contours of $\e_c$ defined in Eq.~\ref{epsilonc} as a function of the curvaton parameters, $\chi_*$, the curvaton VEV, and $m_\chi/\Gamma_\x$ the dimensionless decay time for the cosine curvaton potential, Eq.~(\ref{Vcos}), with $f=10^{16}$~GeV. Curvaton perturbations dominate the primordial scalar power spectrum for $\e_*\gg\e_c$.}
\label{figepscritcos16}
\end{figure}

Figure~\ref{figepscritcos16} shows the parameter $\epsilon_c$ defined in Eq.~(\ref{epsilonc}), which determines the contribution of the curvaton to the scalar power spectrum for a given value of the inflationary energy scale, $H_*$, as a function of $\chi_*$ and $\mc/\Gamma_\x$ for a cosine potential with $f=10^{16}$~GeV. For $\chi_*\ll f$ we recover the previous results for the quadratic potential shown in Figure~\ref{figepscrit}. (Note that the y-axis is linear in Figure~\ref{figepscritcos16} but logarithmic in Figure~\ref{figepscrit}). For values of $\chi_*>f$ the higher-order terms in the potential reduce the potential gradient and hence slow-down the evolution of $\chi$. The curvaton has a larger density when it decays than it would have done for the same initial VEV in a quadratic potential. Thus $\rc$ increases and $\e_c$ decreases relative to the same parameter values in the quadratic potential. In particular this increases the parameter range for which the curvaton dominates the primordial power spectrum, $\e_c>\e_*$, relative to the quadratic case.

\begin{figure}
\centering
\includegraphics[width=0.6\textwidth]{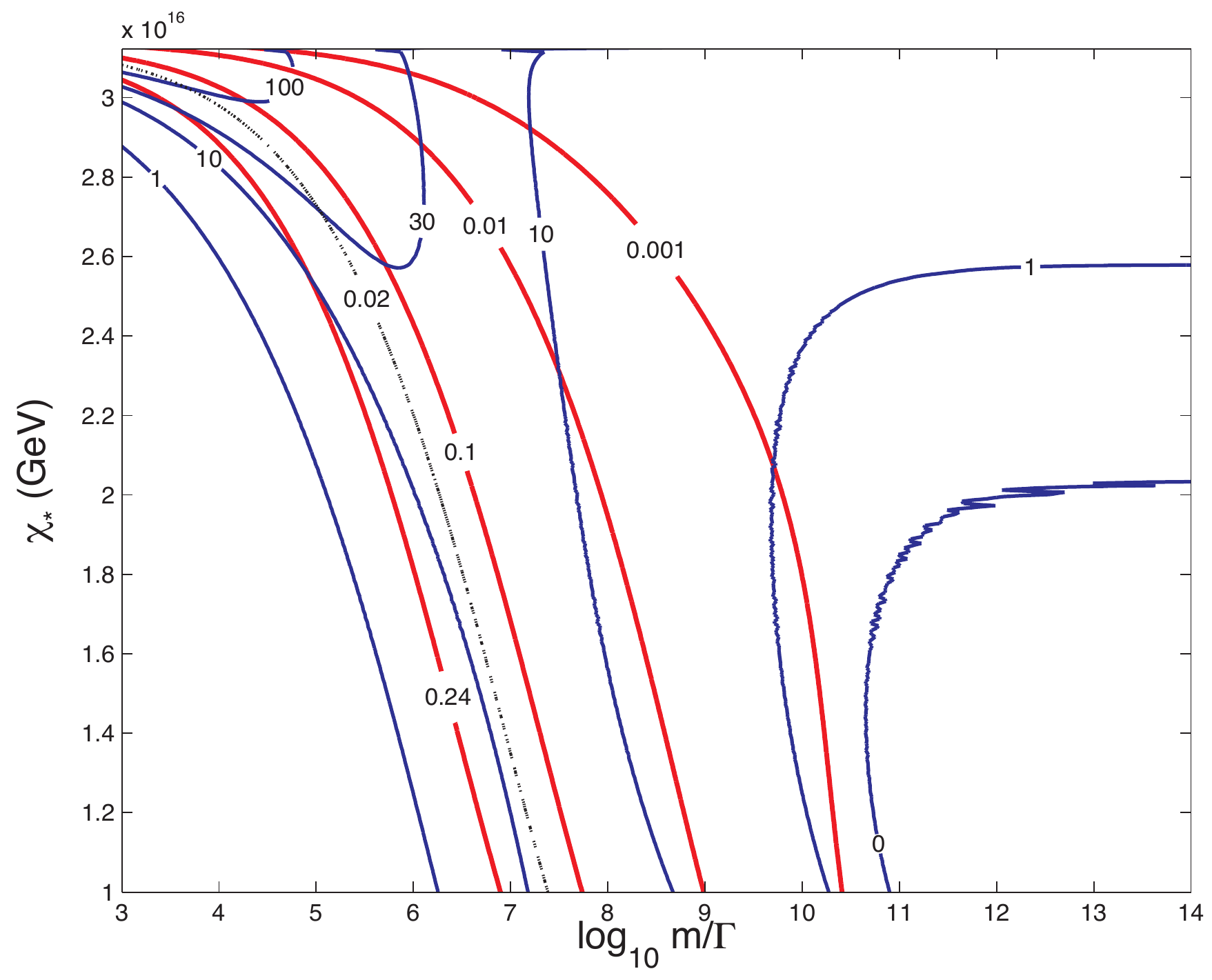}
\caption{The plot show contour lines for the non-linear parameter $\fnl$ (blue lines) and the tensor-to-scalar ratio $\rT$ (thick red lines) as a function of the curvaton parameters, $\chi_*$ and $m_\chi/\Gamma_\x$, for the cosine curvaton potential, Eq.~(\ref{Vcos}), with $f=10^{16}$~GeV, and a fixed value of the inflation slow-roll parameter, $\e_*=0.02$. The black broken line corresponds to $\e_c=0.02$.
%The primordial scalar power spectrum is dominated by inflaton perturbations above and to the left of the thin black dotted line.
}
\label{figfnlrgwcos16eps02}
\end{figure}

In Figure~\ref{figfnlrgwcos16eps02} we show the non-Gaussianity parameter, $\fnl$, and the tensor-scalar ratio, $\rT$, for different curvaton parameter values and a fixed slow-roll parameter, $\e_*=0.02$. Bounds on the tensor-scalar ratio no longer place a lower bound on the decay time, $\mc/\Gamma_\x$, as the tensor-scalar ratio becomes small when $\e_c$ is large for $\chi_*\sim \pi f$ where the curvaton VEV is close to the maximum of the cosine potential. Large positive values of $\fnl$ also become possible for $\chi_*\sim \pi f$ due to the non-linear evolution of the curvaton field, even though $\rc\simeq1$.

For $\chi_*\sim \pi f$ and $\rc\simeq1$ we have from Eq.~(\ref{fnlexpr1})
\be
\fnl \simeq \frac54 \left( \frac{g^{\prime\prime}g}{g^{\prime2}} \right) w_\chi^2 \,,
\ee
and from Eq.~(\ref{gnlexpr1})
\be
\gnl=\frac{25}{24} \left(\frac{g'''g^2}{g'^3}-3\frac{g''g}{g^{\prime2}} \right) \wc^3 \,.
\ee

\subsubsection{Hyperbolic-cosine potential}

We also consider a hyperbolic-cosine potential
\be
 \label{Vcosh}
 V(\chi)= M^4 \left[ \cosh \left( \frac{\chi}{f} \right) - 1\right] \,.
 \ee
For $\chi_*\ll f$ the effective potential  reduces to the quadratic potential (\ref{Vquad}) with $\mc\equiv M^2/f$. Self-interaction terms which become significant for $\chi_*\sim f$ and for $\chi_*\gg f$ the curvaton field becomes massive during inflation and evolves rapidly to smaller values, hence we will assume $\chi_*\lesssim f$ in our discussion.

\begin{figure}
\centering
\includegraphics[width=0.6\textwidth]{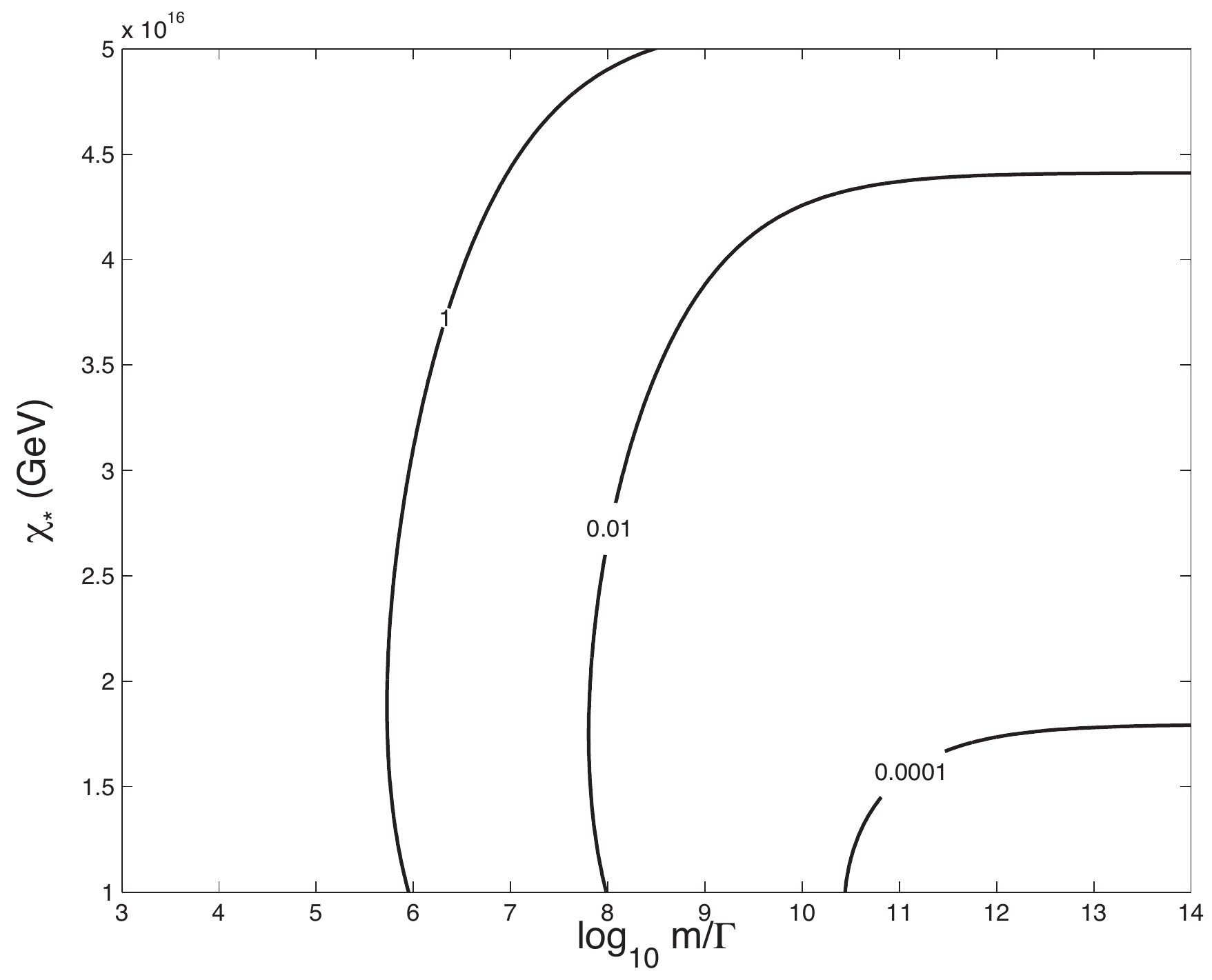}
\caption{The plot show contours of $\e_c$ defined in Eq.~\ref{epsilonc} as a function of the curvaton parameters, $\chi_*$, the curvaton VEV, and $m_\chi/\Gamma$ the dimensionless decay time for the hyperbolic-cosine curvaton potential, Eq.~(\ref{Vcosh}), with $f=10^{16}$~GeV. Curvaton perturbations dominate the primordial scalar power spectrum for $\e_*\gg\e_c$.}
\label{figepscritcosh16}
\end{figure}

We start by studying the relative contribution of the curvaton to the scalar power spectrum characterised by parameter $\epsilon_c$ defined in Eq.~(\ref{epsilonc}).
Figure~\ref{figepscritcosh16} shows $\e_c$ as a function of $\chi_*$ and $\mc/\Gamma_\x$ for a hyperbolic-cosine potential with $f=10^{16}$~GeV. Again, for $\chi_*\ll f$ we recover the previous results for the quadratic potential shown in Figure~\ref{figepscrit}. For values of $\chi_*>f$ the higher-order terms in the potential increase the potential gradient and speed up the evolution of $\chi$ relative to the quadratic potential. The curvaton has a smaller density when it decays than it would have done and thus $\rc$ increases and $\e_c$ decreases relative to the same parameter values in the quadratic potential. This decreases the parameter range for which the curvaton dominates the primordial power spectrum, $\e_c>\e_*$, relative to the quadratic case.

\begin{figure}
\centering
\includegraphics[width=0.6\textwidth]{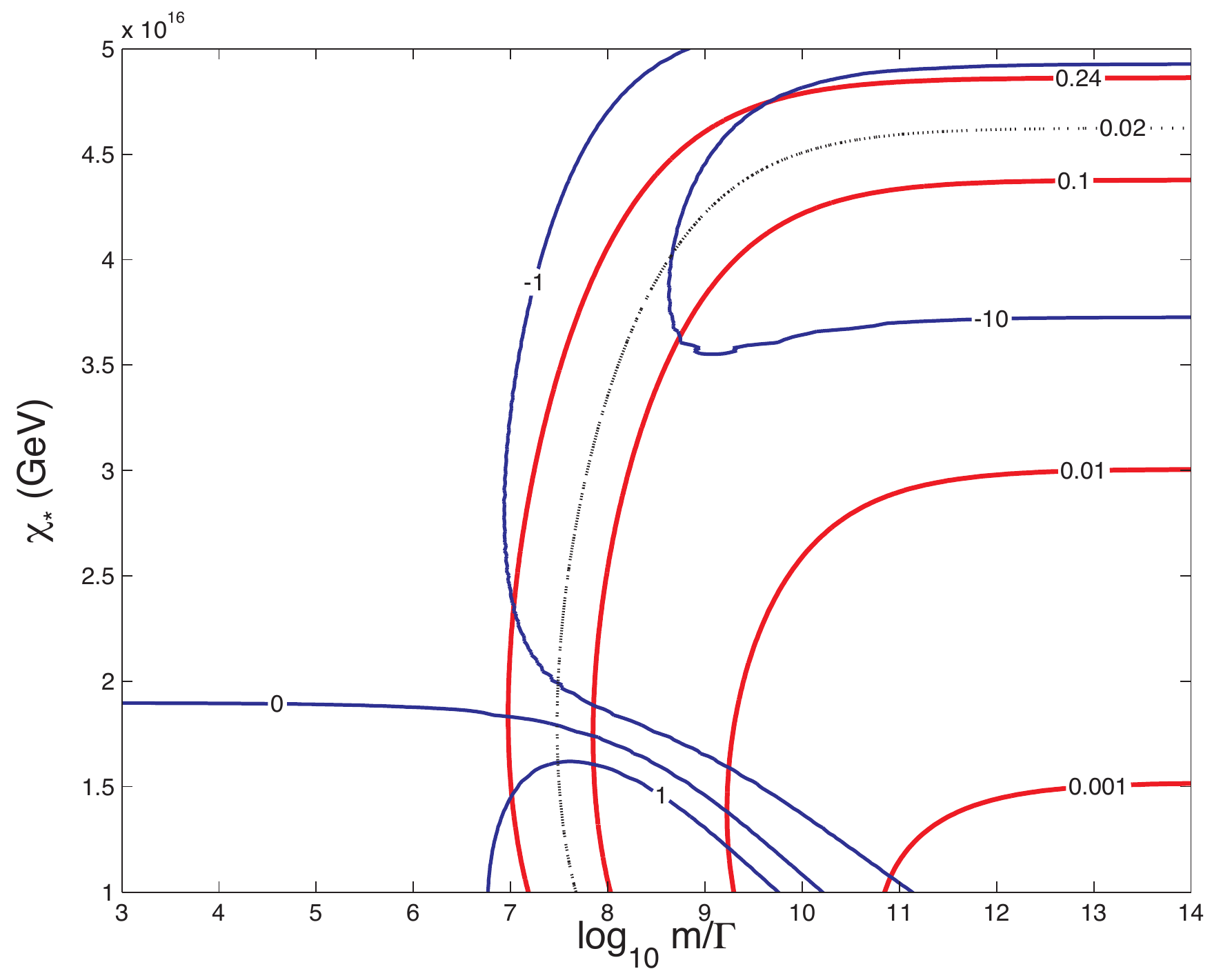}
\caption{The plot show contour lines for the non-linear parameter $\fnl$ (blue lines) and the tensor-to-scalar ratio $\rT$ (thick red lines) as a function of the curvaton parameters, $\chi_*$ and $m_\chi/\Gamma_\x$, for the hyperbolic-cosine curvaton potential, Eq.~(\ref{Vcosh}), with $f=10^{16}$~GeV, and a fixed value of the inflation slow-roll parameter, $\e_*=0.02$. The black broken line corresponds to $\e_c=0.02$.
%The primordial scalar power spectrum is dominated by inflaton perturbations above and to the left of the thin black dotted line.
}
\label{figfnlrgwcosh16eps02}
\end{figure}

In Figure~\ref{figfnlrgwcosh16eps02} we show the non-Gaussianity parameter, $\fnl$, and the tensor-scalar ratio, $\rT$, for different curvaton parameter values and a fixed slow-roll parameter, $\e_*=0.02$. As in the case of a quadratic potential bounds on the tensor-scalar ratio place a lower bound on the decay time, $\mc/\Gamma_\x$ if $\rT<16\e_*$.

Large negative values of $\fnl$ are in principle possible due to the non-linear evolution of the curvaton field for $\chi_*>f$. However, just as in the case of positive $\fnl$ for the quadratic potential, extremely large values are not possible for finite $\e_*$ since $\e_c$ becomes small for $\chi_*\gg f$ and hence $w_\chi\to0$ and $\fnl\to0$ for $\chi_*/f\to+\infty$. The maximum value of $\fnl$ (for sufficiently small decay rates, such that $\rc\simeq1$) occurs when we have $\e_c\sim\e_*$, i.e, at the boundary of curvaton and inflaton limits.

\section{Discussion and Conclusions}

In our previous work \cite{Fonseca:2011iz} we have shown how observables such as the tensor-scalar ratio, $\rT$, and non-linearity parameter, $\fnl$, are related to curvaton model parameters, specifically the curvaton VEV, $\x_*$, and the dimensionless decay rate, $\Gamma_\x/\mc$. In this paper we have allowed for the presence of primordial perturbations due to adiabatic inflaton field fluctuations in addition to isocurvature curvaton field fluctuations during inflation. This introduces an additional model parameter, the slow-roll parameter $\epsilon_*$, which determines the primordial power spectrum due to inflaton field fluctuations relative to the tensor power spectrum. We have constructed an equivalent parameter, $\e_c$, which determines the primordial power spectrum due to curvaton field fluctuations relative to the tensor power spectrum. For $\e_c\ll \e_*$ the curvaton fluctuations dominate the primordial scalar power spectrum, $w_\x\simeq1$, and we recover the results of our previous work \cite{Fonseca:2011iz}. For $\e_c\gg \e_*$ the inflaton fluctuations dominate the primordial scalar power spectrum, $w_\x\ll1$.

In practice we have presented two-dimensional contour plots of the tensor-scalar ratio, $\rT$, and non-linearity parameter, $\fnl$, as functions of $\x_*$ and $\Gamma_\x/\mc$ for fixed values of $\e_*$. We have shown that a curvaton can produced detectable non-Gaussianity and/or gravitational waves for a range of model parameters, even allowing for the presence of inflaton perturbations. For a small slow-roll parameter, $\e_*<\e_c$, very large values of the non-linearity parameters are suppressed ($\fnl\propto (\e_*/\e_c)^2$, $\gnl\propto (\e_*/\e_c)^3$, etc). Nonetheless $\fnl$ may still be produced for $\e_*<\e_c$ when $\rc\ll1$ or in the presence of self-interactions and non-linear curvaton field evolution, $|g^{\prime\prime}g/g^{\prime2}|\gg1$.

To differentiate between different scenarios for the origin of non-Gaussianity we should examine further the statistics of the primordial density field. For example, in the absence of curvaton self-interactions the scale-dependence of the non-linearity parameter is given by \cite{Byrnes:2009pe,Byrnes:2010ft,Byrnes:2010xd,Kobayashi:2012ba}
\be
 n_\fnl \equiv \frac{\partial\ln|\fnl|}{\partial\ln k} = 2 (n_\sigma - 1) - 2 (n_\zeta - 1) = 2 (1-w_\x) (n_\x-n_\p) \,.
\ee
Note that if the curvaton dominates both the power spectrum and the higher-order correlators, $w_\x\simeq 1$, then $\fnl$ is independent of scale.
If the power spectrum is dominated by inflaton perturbations ($w_\x\ll1$), such that $n_\z=n_\phi$, then the bispectrum and higher-order correlators are still dominated by the curvaton perturbations. Hence we generally expect a scale-dependence of the non-linearity parameters $\fnl$ and $\gnl$ that since they determine the higher-order correlators relative to the power spectrum. The higher-order correlators and the power spectrum inherit different scale-dependence from the curvaton and inflaton perturbations respectively. In terms of slow-roll parameters
\be
 n_\fnl \simeq 4 (1-w_\x) (2\e_* + \eta_\x - \eta_\p) \,.
\ee
which is small if we assume slow-roll for both the curvaton and inflaton. However in this case only the inflaton tilt is constrained by current observations of the power spectrum and if the curvaton scale-dependence is large then $n_\fnl$ could be large.

\begin{figure}
\centering
\includegraphics[width=0.6\textwidth]{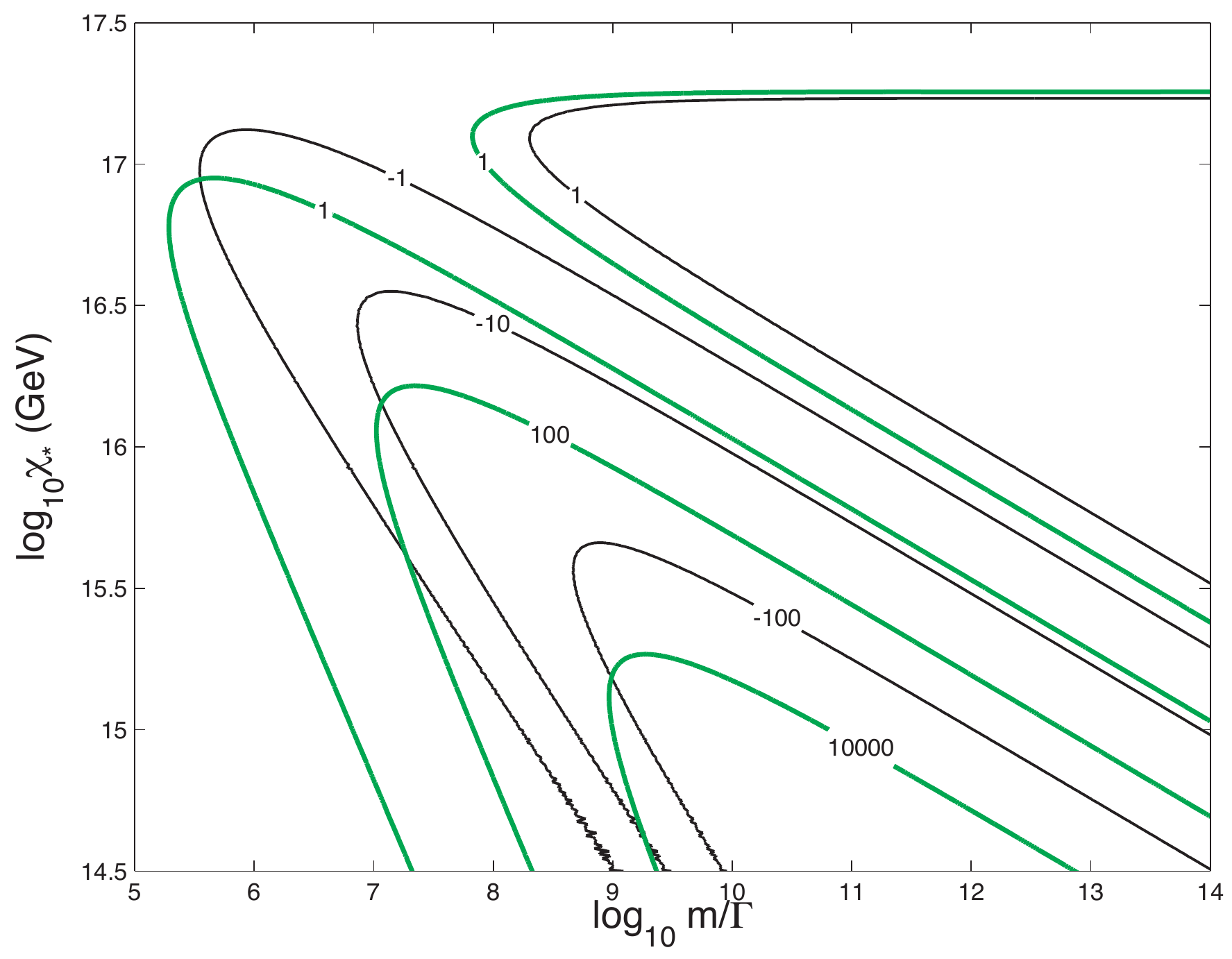}
\caption{Contour lines for the trispectrum non-linear parameters $\gnl$ (thin black lines) and $\tnl$ (thick green lines) as a function of the curvaton parameters, $\chi_*$ and $m_\chi/\Gamma_\x$, for the quadratic curvaton potential, Eq.~(\ref{Vquad}), with a fixed value of the inflation slow-roll parameter, $\e_*=0.02$.
}
\label{figgnltnl}
\end{figure}

The primordial trispectrum also gives important clues about the origin of non-linearity. Figure~\ref{figgnltnl} shows the trispectrum parameters $\gnl$ and $\tnl$ as a function of curvaton parameters for a quadratic curvaton potential. In the absence of self-interactions non-Gaussianity only becomes large when $\rc\ll1$ and in this limit we have
\be
 \label{quadnG}
\gnl \simeq -\frac{10w_\x}{3} \fnl \,, \quad \tnl = \frac{36}{25w_\x} \fnl^2 \,.
\ee
Even allowing for a mixed inflaton+curvaton model with $0\leq w_\x\leq1$, we can eliminate $w_\x$ to obtain a consistency relation between the bispectrum and trispectrum parameters in this case:
\be
 \gnl\tnl \simeq - \frac{24}{5} \fnl^3 \,.
 \ee

Note that from (\ref{quadnG}) we can deduce that $\tnl>0.1\gnl^2$ for a quadratic curvaton potential. If both $\gnl$ and $\tnl$ are large then the curvaton potential must include self-interaction terms \cite{Enqvist:2008gk}. Such self-interactions can give rise to large scale-dependence of $\fnl$ and $\gnl$ even in the curvaton-dominated limit, $w_\x\simeq1$ \cite{Byrnes:2009pe,Byrnes:2010ft,Byrnes:2010xd,Huang:2010cy,Byrnes:2011gh}.

Unlike single-field models of inflation, the predictions of the curvaton model are dependent on the initial value of the curvaton field during inflation. Although it may not be possible to identify a unique initial value for the curvaton, we may be able to specify the expected probability distribution for different models.

If we take a stochastic approach for the distribution of the curvaton VEV \cite{Starobinsky:1994bd} then for a quadratic curvaton potential, and assuming that inflation lasted long enough (and assuming a light curvaton, $\mc^2<H^2$), we find a Gaussian distribution with variance
\be \label{stc}
\langle\x_*^2\rangle=\frac3{8\pi^2}\frac{H_*^4}{\mc^2}\,.
\ee
Huang \cite{Huang:2008ze} has argued that in this case, a detection of primordial non-Gaussianity then places a lower bound on the tensor-scalar ratio.
On the other hand for the self-interacting potential considered here there is an additional scale, $f$, in addition to the effective mass, $\mc$.
For a cosine type potential one obtains an almost uniform distribution for $0\leq\x_*\leq\pi f$ assuming inflation lasts long enough and is at a high-enough energy scale, while for a hyperbolic potential which becomes steep for $\chi_*\gg f$ we expect values with $\x_*\gtrsim f$ to be suppressed.
It would be interesting to interpret observations from future observations of primordial non-Gaussianity and/or tensor-scalar ratio in terms of curvaton model parameters, incorporating a prior probability distribution for the curvaton VEV to marginalise over at least one unknown model parameter.

%In this scenario we assume $\eta_{\x\x}\ll \e_*,\eta_{\p\p}$. For the time being let's take a value for the running, $n_\z=0.973$ \cite{Komatsu:2010fb}, when we marginalize most the running and the tensor-to-scalar ratio.

\acknowledgements{JF was supported by ÒFunda\c c\~{a}o para a Ci\^encia e a Tecnologia (Portugal), fellowships reference number SFRH/BD/40150/2007. DW is supported by STFC grant ST/H002774/1.}

%%%%%%%%%%%%%%%%%%%%%%%%%%%%


\begin{thebibliography}{}
%%%%%%%%%%%%%%%%%%%%%%%%%%%%

 %\cite{Komatsu:2010fb}
\bibitem{Komatsu:2010fb}
  E.~Komatsu {\it et al.}  [WMAP Collaboration],
  %``Seven-Year Wilkinson Microwave Anisotropy Probe (WMAP) Observations:
  %Cosmological Interpretation,''
  arXiv:1001.4538 [astro-ph.CO].
  %%CITATION = ARXIV:1001.4538;%%

%\cite{Mollerach:1989hu}
\bibitem{Mollerach:1989hu}
  S.~Mollerach,
  %``Isocurvature Baryon Perturbations And Inflation,''
  Phys.\ Rev.\ D {\bf 42}, 313 (1990).
  %%CITATION = PHRVA,D42,313;%%

%\cite{Linde:1996gt}
\bibitem{Linde:1996gt}
  A.~D.~Linde and V.~F.~Mukhanov,
  %``Nongaussian isocurvature perturbations from inflation,''
  Phys.\ Rev.\ D {\bf 56}, 535 (1997)
  [astro-ph/9610219].
  %%CITATION = ASTRO-PH/9610219;%%

%\cite{Enqvist:2001zp}
\bibitem{Enqvist:2001zp}
  K.~Enqvist and M.~S.~Sloth,
  %``Adiabatic CMB perturbations in pre - big bang string cosmology,''
  Nucl.\ Phys.\ B {\bf 626}, 395 (2002)
  [hep-ph/0109214].
  %%CITATION = HEP-PH/0109214;%%

     %\cite{}
\bibitem{Lyth:2001nq}
  D.~H.~Lyth and D.~Wands,
  %``Generating the curvature perturbation without an inflaton,''
  Phys.\ Lett.\  B {\bf 524},  5 (2002)
  [arXiv:hep-ph/0110002].
  %%CITATION = PHLTA,B524,5;%%

%\cite{Moroi:2001ct}
\bibitem{Moroi:2001ct}
  T.~Moroi and T.~Takahashi,
  %``Effects of cosmological moduli fields on cosmic microwave background,''
  Phys.\ Lett.\ B {\bf 522}, 215 (2001)
  [Erratum-ibid.\ B {\bf 539}, 303 (2002)]
  [hep-ph/0110096].
  %%CITATION = HEP-PH/0110096;%%

%\cite{Moroi:2002rd}
\bibitem{Moroi:2002rd}
  T.~Moroi and T.~Takahashi,
  %``Cosmic density perturbations from late decaying scalar condensations,''
  Phys.\ Rev.\ D {\bf 66}, 063501 (2002)
  [hep-ph/0206026].
  %%CITATION = HEP-PH/0206026;%%

    %\cite{Lyth:2002my}
\bibitem{Lyth:2002my}
  D.~H.~Lyth, C.~Ungarelli and D.~Wands,
  %``The primordial density perturbation in the curvaton scenario,''
  Phys.\ Rev.\  D {\bf 67}, 023503 (2003)
  [arXiv:astro-ph/0208055].
  %%CITATION = PHRVA,D67,023503;%%

%\cite{Nakayama:2009ce}
\bibitem{Nakayama:2009ce}
  K.~Nakayama and J.~'i.~Yokoyama,
  %``Gravitational Wave Background and Non-Gaussianity as a Probe of the Curvaton Scenario,''
  JCAP {\bf 1001}, 010 (2010)
  [arXiv:0910.0715 [astro-ph.CO]].
  %%CITATION = ARXIV:0910.0715;%%

  %\cite{Fonseca:2011iz}
\bibitem{Fonseca:2011iz}
  J.~Fonseca, D.~Wands,
  %``Non-Gaussianity and Gravitational Waves from Quadratic and Self-interacting Curvaton,''
  Phys.\ Rev.\  {\bf D83}, 064025 (2011).
  [arXiv:1101.1254 [astro-ph.CO]].

%\cite{Bartolo:2002vf}
\bibitem{Bartolo:2002vf}
  N.~Bartolo and A.~R.~Liddle,
  %``The Simplest curvaton model,''
  Phys.\ Rev.\ D {\bf 65}, 121301 (2002)
  [astro-ph/0203076].
  %%CITATION = ASTRO-PH/0203076;%%

%\cite{Ichikawa:2008iq}
\bibitem{Ichikawa:2008iq}
  K.~Ichikawa, T.~Suyama, T.~Takahashi and M.~Yamaguchi,
  %``Non-Gaussianity, Spectral Index and Tensor Modes in Mixed Inflaton and Curvaton Models,''
  Phys.\ Rev.\ D {\bf 78}, 023513 (2008)
  [arXiv:0802.4138 [astro-ph]].
  %%CITATION = ARXIV:0802.4138;%%

  %\cite{Kobayashi:2012ba}
\bibitem{Kobayashi:2012ba}
  T.~Kobayashi and T.~Takahashi,
  %``Runnings in the Curvaton,''
  arXiv:1203.3011 [astro-ph.CO].
  %%CITATION = ARXIV:1203.3011;%%

%\cite{Langlois:2004nn}
\bibitem{Langlois:2004nn}
  D.~Langlois, F.~Vernizzi,
  %``Mixed inflaton and curvaton perturbations,''
  Phys.\ Rev.\  {\bf D70}, 063522 (2004).
  [astro-ph/0403258].

%\cite{Langlois:2008vk}
\bibitem{Langlois:2008vk}
  D.~Langlois, F.~Vernizzi, D.~Wands,
  %``Non-linear isocurvature perturbations and non-Gaussianities,''
  JCAP {\bf 0812}, 004 (2008).
  [arXiv:0809.4646 [astro-ph]].

%\cite{Kinney:2012ik}
\bibitem{Kinney:2012ik} 
  W.~H.~Kinney, A.~M.~Dizgah, B.~A.~Powell and A.~Riotto,
  %``Inflaton or Curvaton? Constraints on Bimodal Primordial Spectra from Mixed Perturbations,''
  arXiv:1203.0693 [astro-ph.CO].
  %%CITATION = ARXIV:1203.0693;%%
  
%\cite{Lyth:2003ip}
\bibitem{Lyth:2003ip}
  D.~H.~Lyth and D.~Wands,
  %``The CDM isocurvature perturbation in the curvaton scenario,''
  Phys.\ Rev.\ D {\bf 68}, 103516 (2003)
  [astro-ph/0306500].
  %%CITATION = ASTRO-PH/0306500;%%

  %\cite{Weinberg:2004kf}
\bibitem{Weinberg:2004kf}
  S.~Weinberg,
  %``Must cosmological perturbations remain non-adiabatic after multi-field inflation?,''
  Phys.\ Rev.\ D {\bf 70}, 083522 (2004)
  [astro-ph/0405397].
  %%CITATION = ASTRO-PH/0405397;%%

%\cite{Malik:2002jb}
\bibitem{Malik:2002jb}
  K.~A.~Malik, D.~Wands and C.~Ungarelli,
  %``Large scale curvature and entropy perturbations for multiple interacting fluids,''
  Phys.\ Rev.\ D {\bf 67}, 063516 (2003)
  [astro-ph/0211602].
  %%CITATION = ASTRO-PH/0211602;%%

  %\cite{Gupta:2003jc}
\bibitem{Gupta:2003jc}
  S.~Gupta, K.~A.~Malik and D.~Wands,
  %``Curvature and isocurvature perturbations in a three-fluid model of curvaton decay,''
  Phys.\ Rev.\ D {\bf 69}, 063513 (2004)
  [astro-ph/0311562].
  %%CITATION = ASTRO-PH/0311562;%%

%\cite{Malik:2006pm}
\bibitem{Malik:2006pm}
  K.~A.~Malik and D.~H.~Lyth,
  %``A numerical study of non-gaussianity in the curvaton scenario,''
  JCAP {\bf 0609}, 008 (2006)
  [astro-ph/0604387].
  %%CITATION = ASTRO-PH/0604387;%%

 %\cite{astro-ph/0607627}
\bibitem{astro-ph/0607627}
  M.~Sasaki, J.~Valiviita and D.~Wands,
  %``Non-Gaussianity of the primordial perturbation in the curvaton model,''
  Phys.\ Rev.\ D\ {\bf 74}, 103003  (2006)
  [astro-ph/0607627].
  %%CITATION = PHRVA,D74,103003;%%

%\cite{Byrnes:2006fr}
\bibitem{Byrnes:2006fr}
  C.~T.~Byrnes and D.~Wands,
  %``Curvature and isocurvature perturbations from two-field inflation in a slow-roll expansion,''
  Phys.\ Rev.\ D {\bf 74}, 043529 (2006)
  [astro-ph/0605679].
  %%CITATION = ASTRO-PH/0605679;%%

%\cite{Gordon:2000hv}
\bibitem{Gordon:2000hv}
  C.~Gordon, D.~Wands, B.~A.~Bassett and R.~Maartens,
  %``Adiabatic and entropy perturbations from inflation,''
  Phys.\ Rev.\ D {\bf 63}, 023506 (2001)
  [astro-ph/0009131].
  %%CITATION = ASTRO-PH/0009131;%%

%\cite{Wands:2002bn}
\bibitem{Wands:2002bn}
  D.~Wands, N.~Bartolo, S.~Matarrese and A.~Riotto,
  %``An observational test of two-field inflation,''
  Phys.\ Rev.\  D {\bf 66}, 043520 (2002)
  [arXiv:astro-ph/0205253].
  %%CITATION = PHRVA,D66,043520;%%

%\cite{Dimopoulos:2011gb}
\bibitem{Dimopoulos:2011gb}
  K.~Dimopoulos, K.~Kohri, D.~H.~Lyth and T.~Matsuda,
  %``The inflating curvaton,''
  arXiv:1110.2951 [astro-ph.CO].
  %%CITATION = ARXIV:1110.2951;%%

%\cite{Lyth:2004gb}
\bibitem{Lyth:2004gb}
  D.~H.~Lyth, K.~A.~Malik and M.~Sasaki,
  %``A General proof of the conservation of the curvature perturbation,''
  JCAP {\bf 0505}, 004 (2005)
  [astro-ph/0411220].
  %%CITATION = ASTRO-PH/0411220;%%

%\cite{Lyth:2009zz}
\bibitem{Lyth:2009zz}
  D.~H.~Lyth, A.~R.~Liddle,
  %``The primordial density perturbation: Cosmology, inflation and the origin of structure,''
  Cambridge, UK: Cambridge Univ. Pr. (2009) 497 p.

%\cite{Wands:2003pw}
\bibitem{Wands:2003pw}
  D.~Wands,
  %``Inflationary parameters and primordial perturbation spectra,''
  New Astron.\ Rev.\  {\bf 47}, 781 (2003)
  [arXiv:astro-ph/0306523].
  %%CITATION = ASTRE,47,781;%%

%\cite{Lyth:2005fi}
\bibitem{Lyth:2005fi}
  D.~H.~Lyth, Y.~Rodriguez,
  %``The Inflationary prediction for primordial non-Gaussianity,''
  Phys.\ Rev.\ Lett.\  {\bf 95}, 121302 (2005).
  [astro-ph/0504045].

 %\cite{Byrnes:2006vq}
\bibitem{Byrnes:2006vq}
  C.~T.~Byrnes, M.~Sasaki and D.~Wands,
  %``The primordial trispectrum from inflation,''
  Phys.\ Rev.\ D {\bf 74}, 123519 (2006)
  [astro-ph/0611075].
  %%CITATION = ASTRO-PH/0611075;%%

%\cite{Dimopoulos:2003az}
\bibitem{Dimopoulos:2003az}
  K.~Dimopoulos, D.~H.~Lyth, A.~Notari and A.~Riotto,
  %``The Curvaton as a pseudoNambu-Goldstone boson,''
  JHEP {\bf 0307}, 053 (2003)
  [hep-ph/0304050].
  %%CITATION = HEP-PH/0304050;%%

%\cite{Kawasaki:2008mc}
\bibitem{Kawasaki:2008mc}
  M.~Kawasaki, K.~Nakayama and F.~Takahashi,
  %``Hilltop Non-Gaussianity,''
  JCAP {\bf 0901}, 026 (2009)
  [arXiv:0810.1585 [hep-ph]].
  %%CITATION = ARXIV:0810.1585;%%

%\cite{Chingangbam:2009xi}
\bibitem{Chingangbam:2009xi}
  P.~Chingangbam and Q.~-G.~Huang,
  %``The Curvature Perturbation in the Axion-type Curvaton Model,''
  JCAP {\bf 0904}, 031 (2009)
  [arXiv:0902.2619 [astro-ph.CO]].
  %%CITATION = ARXIV:0902.2619;%%

%\cite{Suyama:2007bg}
\bibitem{Suyama:2007bg}
  T.~Suyama and M.~Yamaguchi,
  %``Non-Gaussianity in the modulated reheating scenario,''
  Phys.\ Rev.\ D {\bf 77}, 023505 (2008)
  [arXiv:0709.2545 [astro-ph]].
  %%CITATION = ARXIV:0709.2545;%%

%\cite{Enqvist:2008gk}
\bibitem{Enqvist:2008gk}
  K.~Enqvist and T.~Takahashi,
  %``Signatures of Non-Gaussianity in the Curvaton Model,''
  JCAP {\bf 0809}, 012 (2008)
  [arXiv:0807.3069 [astro-ph]].
  %%CITATION = ARXIV:0807.3069;%%

  %\cite{Byrnes:2009pe}
\bibitem{Byrnes:2009pe}
  C.~T.~Byrnes, S.~Nurmi, G.~Tasinato and D.~Wands,
  %``Scale dependence of local f_NL,''
  JCAP {\bf 1002}, 034 (2010)
  [arXiv:0911.2780 [astro-ph.CO]].
  %%CITATION = ARXIV:0911.2780;%%

%\cite{Byrnes:2010ft}
\bibitem{Byrnes:2010ft}
  C.~T.~Byrnes, M.~Gerstenlauer, S.~Nurmi, G.~Tasinato and D.~Wands,
  %``Scale-dependent non-Gaussianity probes inflationary physics,''
  JCAP {\bf 1010}, 004 (2010)
  [arXiv:1007.4277 [astro-ph.CO]].
  %%CITATION = ARXIV:1007.4277;%%

%\cite{Byrnes:2010xd}
\bibitem{Byrnes:2010xd}
  C.~T.~Byrnes, K.~Enqvist and T.~Takahashi,
  %``Scale-dependence of Non-Gaussianity in the Curvaton Model,''
  JCAP {\bf 1009}, 026 (2010)
  [arXiv:1007.5148 [astro-ph.CO]].
  %%CITATION = ARXIV:1007.5148;%%

  %\cite{Huang:2010cy}
\bibitem{Huang:2010cy}
  Q.~-G.~Huang,
  %``Negative spectral index of $f_{NL}$ in the axion-type curvaton model,''
  JCAP {\bf 1011}, 026 (2010)
  [Erratum-ibid.\  {\bf 1102}, E01 (2011)]
  [arXiv:1008.2641 [astro-ph.CO]].
  %%CITATION = ARXIV:1008.2641;%%

%\cite{Byrnes:2011gh}
\bibitem{Byrnes:2011gh}
  C.~T.~Byrnes, K.~Enqvist, S.~Nurmi and T.~Takahashi,
  %``Strongly scale-dependent polyspectra from curvaton self-interactions,''
  JCAP {\bf 1111}, 011 (2011)
  [arXiv:1108.2708 [astro-ph.CO]].
  %%CITATION = ARXIV:1108.2708;%%

  %\cite{Starobinsky:1994bd}
\bibitem{Starobinsky:1994bd}
  A.~A.~Starobinsky and J.~Yokoyama,
  %``Equilibrium state of a selfinteracting scalar field in the De Sitter background,''
  Phys.\ Rev.\ D {\bf 50}, 6357 (1994)
  [astro-ph/9407016].
  %%CITATION = ASTRO-PH/9407016;%%

%\cite{Huang:2008ze}
\bibitem{Huang:2008ze}
  Q.~-G.~Huang,
  %``Large Non-Gaussianity Implication for Curvaton Scenario,''
  Phys.\ Lett.\ B {\bf 669}, 260 (2008)
  [arXiv:0801.0467 [hep-th]].
  %%CITATION = ARXIV:0801.0467;%%

\end{thebibliography}
\end{document}